\documentclass[12pt, oneside]{report} 
\usepackage[utf8]{inputenc}
\usepackage[english]{babel}
\usepackage{csquotes}
\usepackage[usenames,dvipsnames]{color}
\usepackage{graphicx}
\usepackage[export]{adjustbox}   
\usepackage{multicol,caption}
\usepackage{amsmath}
\usepackage{amssymb}
\usepackage{gensymb}
\usepackage[]{hyphsubst} 
\usepackage[parfill]{parskip}
\usepackage{multirow} 
\usepackage{supertabular} 
\usepackage{tabularx}
\usepackage{mathrsfs} 
\usepackage{trfsigns} 
\usepackage{pdfpages}
\usepackage{flafter} 
\usepackage{isotope} 
\usepackage{makecell}

\usepackage{setspace}

\DeclareUnicodeCharacter{3B2}{\ss}  

\usepackage[style=ieee, backend=biber, maxnames=2, minnames=1]{biblatex}

\usepackage{ fancyhdr }

\usepackage[hidelinks,pdfusetitle,pdfsubject={MHD Ion Mass Effects Simulations for Tokamaks},pdfkeywords={MHD, Tokamak, plasma physics, ELM, peeling-ballooning, instability},pdfauthor={Matthias Rosenthal}]{hyperref}

\usepackage[acronym]{glossaries}

\makeatletter
\patchcmd{\lp@dimens}
  {\epsfig{figure=\lp@epsfile}}
  {\includegraphics{\lp@epsfile}}
  {}{}
\patchcmd{\lp@dimens}
  {\epsfig{figure=\lp@epsfile, width=\lp@tmpx, height=\lp@tmpy}}
  {\includegraphics[width=\lp@tmpx,height=\lp@tmpy]{\lp@epsfile}}
  {}{}
\makeatother

\usepackage{geometry}
\newcommand{\clearemptydoublepage} {
  \newpage{
  	 \pagestyle{empty}
	 \cleardoublepage
  }
}

\bibliography{bachelors_thesis_tum.bib}

\makeglossaries
\newacronym{elm}{ELM}{Edge-Localized Mode}
\newacronym{lcfs}{LCFS}{last closed flux surface}
\newacronym{mhd}{MHD}{Magnetohydrodynamics}
\newacronym{aug}{AUG}{ASDEX Upgrade}
\newacronym{lfs}{LFS}{Low-field side}
\newacronym{hfs}{HFS}{High-field side}
\newacronym{jet}{JET}{Joint European Torus}
\newacronym{sol}{SOL}{Scrape-off layer}

\begin{document}

\title{Magneto-Hydrodynamic Simulations of Pedestal Instabilities for Tokamak Plasmas with Different Ion Masses}
\author{Matthias Rosenthal}

\begin{spacing}{1.15}

\pagenumbering{alph}
    \begin{titlepage}
    \includegraphics[width=0.6\textwidth, valign=t]{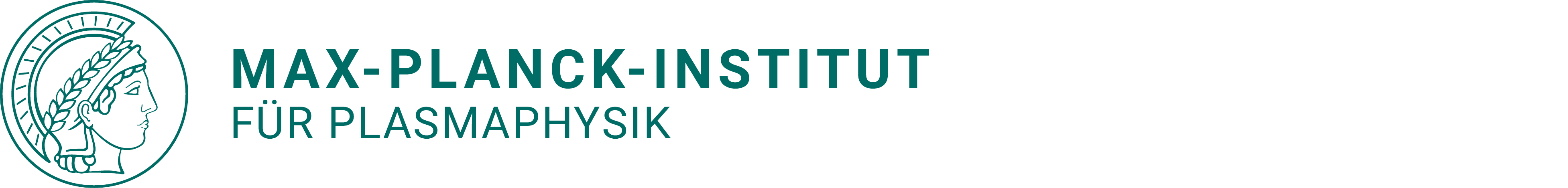}
    \hfill
    \includegraphics[width=0.05\textwidth, valign=t]{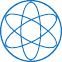}
    \includegraphics[width=0.1\textwidth, valign=t]{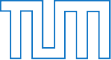}

	\begin{center}
        \vspace{3cm}
		{\par Bachelor's thesis in physics}
		\vspace{0.5cm}
		{\par \large\bfseries Magneto-Hydrodynamic Simulations of Pedestal Instabilities for\newline To\-ka\-mak Plasmas with Different Ion Masses}
			\vspace{0.5cm}\\
		{Matthias Rosenthal\\ Matriculation number: 03765569}
		\vspace{0.2cm}
		{\par \today}
		\vspace{0.2cm}
		{\par School of natural sciences
		\\ Physics department}
			\vfill
		{\large \par}
	\end{center}

	\setlength\tabcolsep{0pt}  
  
    \begin{tabular}{ p{0.5\textwidth} p{0.5\textwidth} }
     Supervisor:
       &
       Prof. Dr. Sibylle Günter
     \\
     Additional supervisors:
       &
       Dr. Matthias Hölzl
     \\
     
       &
       Dr. Andres Cathey
     \\
    \end{tabular}
\end{titlepage}

\clearemptydoublepage
\tableofcontents

\clearemptydoublepage
\listoffigures
\addcontentsline{toc}{chapter}{\listfigurename}  

\clearemptydoublepage
\listoftables
\addcontentsline{toc}{chapter}{\listtablename}  

\clearemptydoublepage
\printglossary[nonumberlist, type=\acronymtype, nopostdot, nogroupskip,title=Acronyms]
\addcontentsline{toc}{chapter}{Table of Acronyms}  
\clearemptydoublepage

\pagenumbering{arabic}

\clearemptydoublepage
\chapter{Introduction}
\par Nuclear fusion is one of the upcoming options to generate fossil-free energy, and is currently under development. At the present time, the most feasible way of doing this is to fuse deuterium (\isotope[2][]{H}) and tritium (\isotope[3][]{H}) nuclei into an energetic neutron and an energetic helium nucleus. To achieve a reasonable cross-section, the product of density, temperature, and energy confinement time needs to be sufficiently high. One way of achieving this is by magnetic confinement.

\par At the required temperatures of more than $100 \cdot 10^6$\,K, electrons separate from their corresponding nuclei, which leads to a distinct state of matter called plasma. One way of confining this plasma magnetically is by means of a tokamak reactor, in which the plasma forms a torus.

\par
\begin{figure}
  \center{\includegraphics[width=1.0\textwidth]{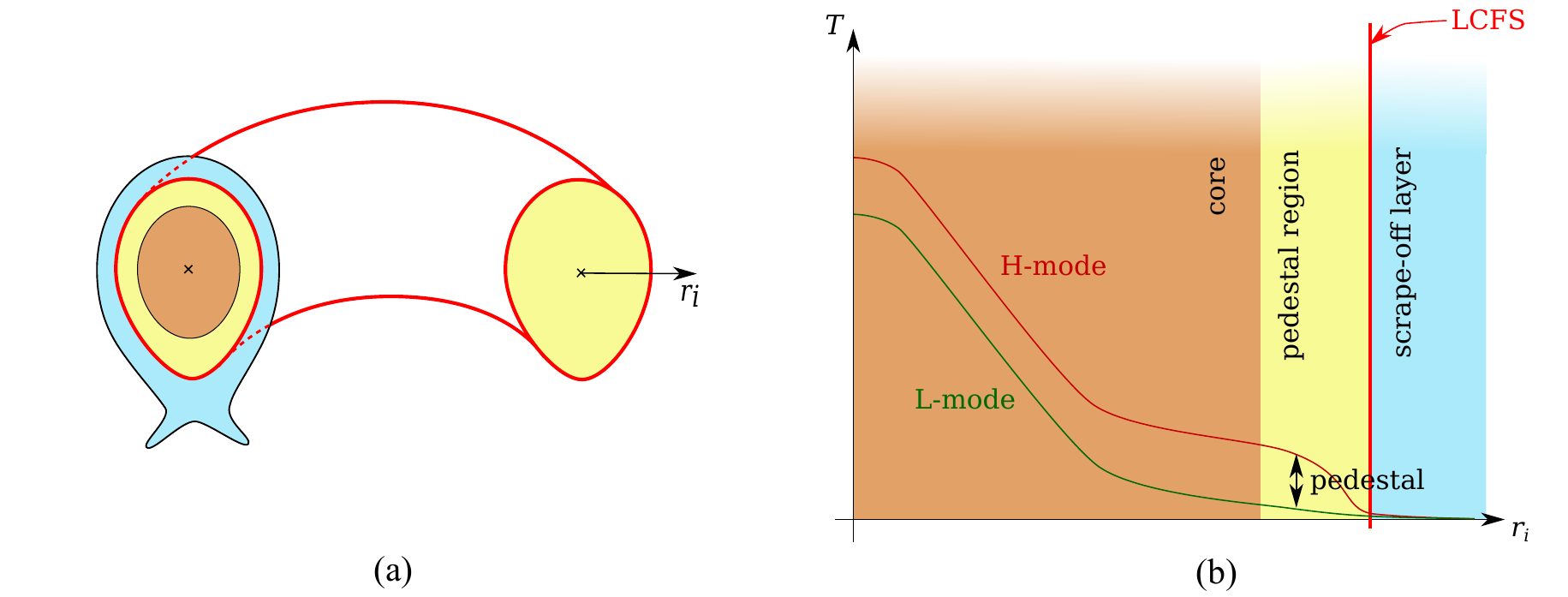}}
  \captionof{figure}{Different sections of a tokamak plasma in H-mode and comparison to L-mode}
  \label{fig:temperature_parts_and_l_vs_h}
\end{figure}

 The common temperature, pressure and density profiles that develop under moderate conditions in a tokamak are summarized under the state of L-mode. All of those three quantities typically increase towards the plasma core (where $r_i = 0$), as indicated in figure \ref{fig:temperature_parts_and_l_vs_h}. When injecting more power into the plasma, a so-called pedestal builds up at the plasma edge. The whole profile is lifted by this pedestal, which is the reason for naming it like that. This is also shown in figure \ref{fig:temperature_parts_and_l_vs_h}.

\par The profiles in this high-confinement mode (H-mode) can therefore be split into three sections, as shown in figure \ref{fig:temperature_parts_and_l_vs_h}. The core area is lifted by the pedestal. The pedestal region itself is the area of steep gradient at the edge. The confined plasma ends at the last closed flux surface (\acrshort{lcfs}), indicated in red in the aforementioned figure. The scrape-off layer is the whole area at the outside of the plasma.

\par Although the H-mode enables to reach parameters closer to fusion conditions, it comes at the downside of the excitation of edge-localized modes (\acrshort{elm}s). These plasma instabilities occur repetitively upon entering this high confinement regime. During each \acrshort{elm} event, a substantial fraction of the plasma material and thermal energy content is expelled, and thus hot plasma is incident on the material walls of the tokamak. For fusion power plant relevant conditions (like ITER, for instance), the damage caused by ELMs on the plasma facing components is expected to be beyond acceptable limits \cite{eichELMDivertorPeak2017}. As the name indicates, \acrshort{elm}s appear at the edge in the pedestal which is in the region indicated in yellow in figure \ref{fig:temperature_parts_and_l_vs_h}. They are the result of coupled peeling-ballooning modes \cite{wilsonTheoryExplosiveIdeal2004}.

\par Many current experiments like ASDEX Upgrade do not study plasmas of the isotopes deuterium and tritium (called D-T) which are required for fusion, but instead work with protium (\isotope[1][]{H}), deuterium and helium only, because this is much easier to carry out. Experimentally, differences in the results where found when using different isotopes of hydrogen. One such difference is the increase in power threshold that is required to move from L-mode to H-mode, which was shown to result mainly from physics happening at the plasma core \cite{callahanPowerThreshold2023}, and also the plasma pedestal more generally \cite{frassinettiPedestalStructureStability2021}.

\par Instead, this thesis aims to improve the understanding of the instabilities that are the origin of \acrshort{elm}s by means of physics occurring at the pedestal. This is done by simulating an H-mode plasma configuration in the medium size ASDEX upgrade (\acrshort{aug}) tokamak device, using the JOREK code described in section \ref{approach_jorek}. In contrast to experimental studies, the simulation makes it possible to avoid changes caused by effects which are not directly contributing to \acrshort{elm}s and their precursors. Thus it is possible in the simulations to separate the ion mass effect from other differences the corresponding experiments may have (e.g., due to different heat source distributions with different isotopes, or due to different core turbulent transport). Therefore, it is also possible to make at least a qualitative statement about which experimentally observed changes are due to magnetohydrodynamic (\acrshort{mhd}) effects.

\par This thesis is organized as follows: in chapter \ref{fundamentals}, a review of plasma physics with focus on tokamak reactors is given. In particular, section \ref{different_isotopes_fundamentals} lists some known experimental differences between ion masses. In chapter \ref{approach}, the JOREK framework and the specific \acrshort{mhd} model used here is introduced. Chapter \ref{results} is about the parameters and results of the simulations that compare the different ion masses. Here, in section \ref{normalization}, the comparative simulation setup is discussed. In section \ref{single_n_simulation}, the results of simulations for only one toroidal mode are presented and the differences are investigated. In section \ref{multi_n_simulation}, a first simulation is performed with multiple toroidal harmonics, reaching into the violent non-linear phase of an \acrshort{elm}-like event. This is to assess the order of magnitude, that the difference between the ion masses coming from MHD have on the losses occurring during an \acrshort{elm} crash.

\clearemptydoublepage
\chapter{Fundamentals}
\label{fundamentals}

\section{Plasma Physics}
\label{plasma_physics}

\par In a magnetic field, charged particles with non-zero velocity parallel to $\vec{B}$ travel along the field lines, while gyrating around the field lines with the gyro radius $r_L = \frac{m v_\perp}{q B}$. Here, $m$, $q$ and $v_\perp$ are the particle's mass, charge, and velocity component orthogonal to the magnetic field direction, respectively. In a plasma, electrons and ions have a very high temperature\footnote{In the context of magnetically confined plasmas, temperatures in the order of 1\,eV are referred to as low temperatures, whereas high temperatures are in the order of 1 to 10\,keV.}, which corresponds to a high thermal velocity. The thermal velocity is distributed on the velocity parallel to the magnetic field lines $v_\parallel$, and the velocity $v_\perp$ perpendicular to them. As visible from the formula for $r_L$, this leads to a small gyro radius below one centimeter for magnetic fields in the order of a tesla, which is typical for fusion reactors \cite[113]{freidbergPlasmaPhysicsFusion2007}. Therefore, when no external forces are present and collisions with other particles are neglected, both ions and electrons are effectively trapped along the magnetic field lines.

\par Typical fusion plasma parameters mentioned in \cite[111]{freidbergPlasmaPhysicsFusion2007} are\footnote{In plasma physics contexts, temperatures are expressed in terms of the corresponding energy. The conversion is done by $T_\mathrm{plasma\ physics} = k_B \, T_\mathrm{Kelvin}$ \cite{mohrCODATAEnergyConversion2024}. 
} $T = 15\,\mathrm{keV} = 174 \cdot 10^6\,\mathrm{K}$, $p = 7$\,bar, which leads to a density of $n = \frac{p}{T} = 2.9\cdot 10^{20}\,\frac{1}{\mathrm{m}^3}$.

\subsection{MHD and Force Balance of a Confined Plasma}
\label{force_balance_of_plasma}

\par In this thesis, the JOREK code is used for plasma simulations, which solves extended visco-resistive \acrshort{mhd} equations with time evolution equations for the magnetic field, electric field, density, momentum and pressure. This will be discussed further in chapter \ref{approach}. In this section, a simplified form of the \acrshort{mhd} is presented instead, to illustrate basic principles governing a tokamak H-mode plasma.
The introduction follows roughly chapter 2 from \cite{militelloBoundaryPlasmaPhysics2022}.

\par A magnetically confined plasma consists of ions and electrons. The electrons, as well as each ionization state of the ions, can be represented by a particle species labelled with $\alpha$. Their behavior can be described by means of a distribution function $f_\alpha(\vec{x}, \vec{v})$, which at a given time point $t$ represents the probability density of finding a particle of species $\alpha$ at location $\vec{x}$ with velocity $\vec{v}$. 
The time evolution of the value of such a distribution at location $\vec{x}$ and velocity $\vec{v}$ is governed by a system of an infinite number of equations called BBGKY, which is derived from the Liouville equation of statistical thermodynamics. Under the assumption that only binary collisions are relevant, this yields the following equation for the time evolution, with electric and magnetic fields $\vec{E}$ and $\vec{B}$ at location $\vec{x}$ \cite[32]{militelloBoundaryPlasmaPhysics2022}: 

\begin{equation}
\label{eqn:kinetic_boltzmann_equation}
\frac{\partial f_\alpha}{\partial t} + \vec{v} \cdot \frac{\partial f_\alpha}{\partial \vec{x}} + \frac{q_\alpha ( \vec{E} + \vec{v} \times \vec{B})}{m_\alpha} \cdot \frac{\partial f_\alpha}{\partial \vec{v}} = \sum_\beta C_{\alpha \beta}(f_\alpha, f_\beta)
\end{equation}

Here, $q_\alpha$ is the charge and $m_\alpha$ is the mass of a particle of species $\alpha$, while $C_{\alpha \beta}$ is describing the collisions and is to be further specified. Additionally to this equation, the Maxwell equations are used to link $\vec{E}$ and $\vec{B}$ to the charge density ${\rho = \sum_\alpha q_\alpha n_\alpha}$ and current ${\vec{J} = \sum_\alpha q_\alpha n_\alpha \vec{u}_\alpha}$. Here, the density ${n_\alpha = \int \mathrm{d}\vec{v} f_\alpha}$ and average velocity ${\vec{u}_\alpha = \int \mathrm{d} \vec{v} \, \vec{v} f_\alpha}$ were introduced. They are examples of so-called moments, which are defined as ${w_\alpha = \int \mathrm{d}\vec{v} G(\vec{v}) f_\alpha}$, where $G$ and $w_\alpha$ are both tensors of the same rank (e.g., $G(\vec{v}) = 1$ for $n$). It is shown in \cite[section 2.3]{militelloBoundaryPlasmaPhysics2022} how the collisions between ions and electrons can be incorporated into $C_{\alpha \beta}$, which drives the system to a state described by a Maxwell-Boltzmann distribution (regarding the velocities). By integrating the product of equation (\ref{eqn:kinetic_boltzmann_equation}) and a tensor $G(\vec{v})$ over the velocity space, one can then arrive at an equation for the time evolution of $n_\alpha$ (using ${G(\vec{v}) = 1}$), $\vec{u}_\alpha$ (using ${G(\vec{v}) = m_\alpha \vec{v}}$), as well as an equation governing the energy (using ${G(\vec{v}) = 1/2 m_\alpha \vec{v}^2}$). An issue of these equations is, that they always depend on moments of one order higher: the equation for $n_\alpha$ depends on $\vec{u}_\alpha$, the equation for $\vec{u}_\alpha$ depends on momentums with $G \propto \vec{v}^2$, and so on, forming a hierarchy of an infinite number of equations governing the evolution of each moment. This can be resolved by so-called closures, which provide assumptions / approximations that link higher-order momentums, for which no explicit evolution equation is to be formulated, to lower-order momentums.

\par One such closure is called ideal \acrshort{mhd}. The most important requirements are, that the characteristic timescale is longer than the mean time between collisions and that the typical length scale of variations is longer than the mean free path \cite[58]{militelloBoundaryPlasmaPhysics2022}. Also, the typical velocity of the analyzed phenomena must be comparable to the ion thermal velocity of the plasma, $u \sim v_{t, i}$ \cite[67]{militelloBoundaryPlasmaPhysics2022}. It should be noted, that the gyro radius is much smaller than the mean free path \cite[59]{militelloBoundaryPlasmaPhysics2022}. Only ions and electrons are regarded, and their profiles are combined to the \acrshort{mhd} temperature $T = T_e + T_i$ and pressure $p = p_e + p_i$. Also, the macroscopic velocity is defined as $\vec{v} = (\rho_e \vec{v}_e + \rho_i \vec{v}_i) / (\rho_e + \rho_i)$ (note that here, the same symbol $\vec{v}$ is used that was previously used for the phase space coordinate). The macroscopic velocity is approximated as $\vec{v} \approx \vec{v}_i$, because the ion's momentum is much larger than the electron's momentum due to the higher ion mass $m_i \gg m_e$. The resulting equations describe the plasma as a single fluid, which is why it is also possible to start their derivation by splitting the plasma into small fluid elements, as done in chapter 10 of \cite{freidbergPlasmaPhysicsFusion2007}. Also, quasi-neutrality is assumed, leading to\footnote{Charge densities can still be modelled, but in a confined plasma, larger charge densities are quickly balanced out. Nevertheless, the resulting electric fields are important for the overall behavior of the confined plasma, as shown below \cite[55]{militelloBoundaryPlasmaPhysics2022}.} ${n = n_e = n_i}$.
This also leads to a current $\vec{J} = n \, e (\vec{v}_i - \vec{v}_e)$, with $e$ being the Coulomb charge\footnote{Here, singly charged ions are assumed, but in principle, ideal \acrshort{mhd} can also be applied to ions with higher charge number.}.
It is important to note that the fluid element that is under consideration is moving with $\vec{v}$: All equations are expressed in the so-called \acrshort{mhd} reference frame, that is, within a reference frame that moves with $\vec{v}$. 
One of the resulting equations is the so-called momentum equation:

\begin{equation}
\label{eqn:basic_mhd}
\rho \frac{d \vec{v}}{d t} = \vec{J} \times \vec{B} - \vec{\nabla} p
\end{equation}

\par In stationary equilibrium, equation \ref{eqn:basic_mhd} can be used with $d \vec{v} / dt = 0$. This leads to the following equation: 

\begin{equation}
\label{eqn:mhd_force_balance}
0 = \vec{J} \times \vec{B} -\vec{\nabla}p
\end{equation}

\par
\begin{figure}
  \center{\includegraphics[width=0.7\textwidth]{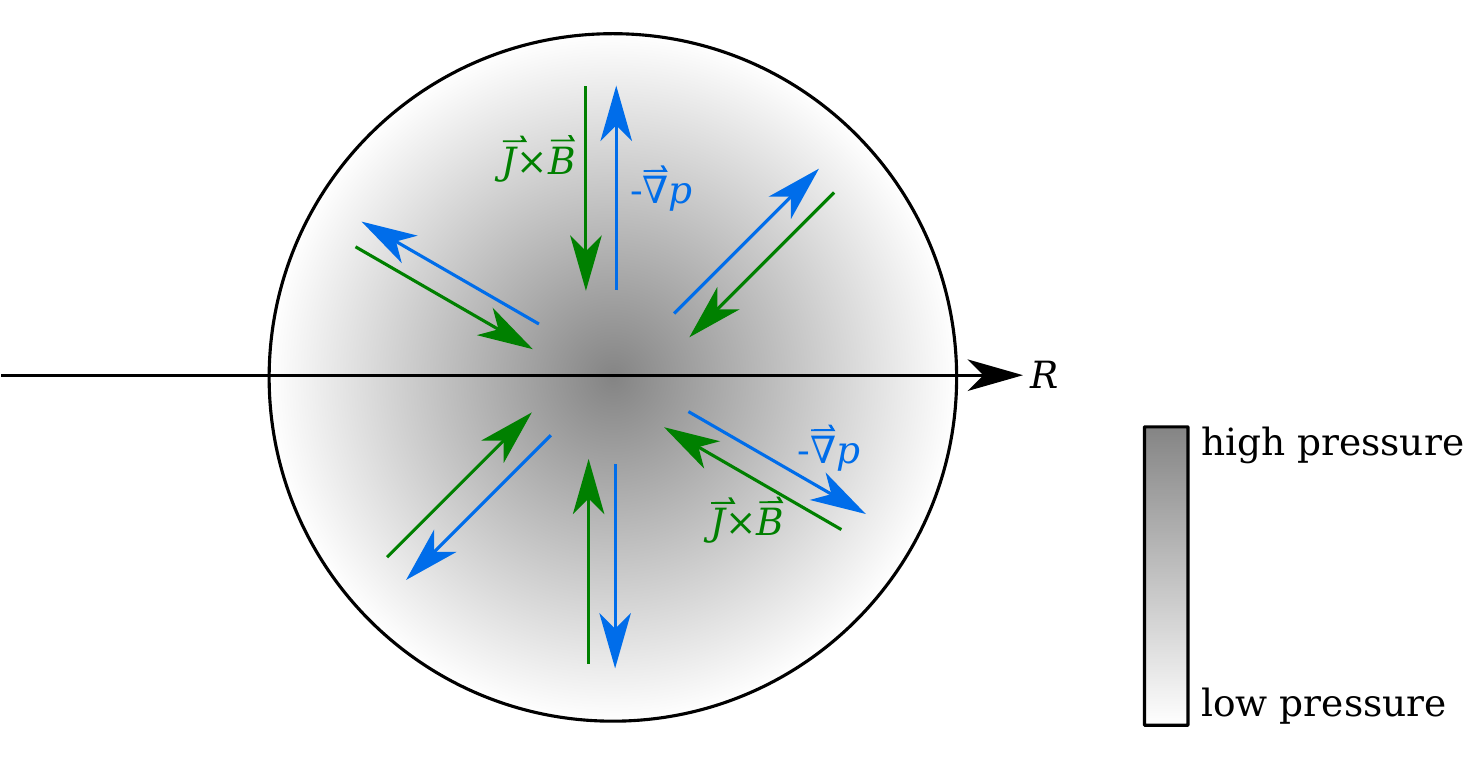}}
  \captionof{figure}{Basic magnetohydrodynamic force balance of a confined plasma}
  \label{fig:plasma_pressure_force}
\end{figure}

 This can be understood as a stationary force balance, as illustrated in figure \ref{fig:plasma_pressure_force}.

\par Using Ampère's law, $\mu_0 \vec{J} = \vec{\nabla} \times \vec{B}$, one can split the $\vec{J} \times \vec{B}$ term as follows:

\begin{equation}
\label{eqn:plasma_pressure}
\vec{J} \times \vec{B} = \frac{(\vec{B} \cdot \vec{\nabla})\vec{B}}{\mu_0} - \vec{\nabla} \left( \frac{\vec{B}^2}{2 \mu_0} \right) \mathrm{,}
\end{equation}

\par With the unit vector $\vec{b} = \vec{B} / |\vec{B}|$ indicating the magnetic field's direction, one can define the curvature vector of the magnetic field $\vec{\kappa} = (\vec{b} \cdot \vec{\nabla}) \vec{b}$ (inversely proportional to $r$ in case of an axisymmetric reactor). Then, equation \ref{eqn:plasma_pressure} can be rewritten to also include the magnetic tension force $\vec{F}_T$ \cite[263]{freidbergPlasmaPhysicsFusion2007}: 

\begin{equation}
\vec{J} \times \vec{B} = \underbrace{\frac{B^2}{\mu_0} \vec{\kappa}}_{\vec{F}_T} - \vec{\nabla}_\perp \left( \frac{B^2}{2 \mu_0} \right)
\end{equation}

\par The right term is called the magnetic pressure force, with the magnetic pressure $\frac{B^2}{2 \mu_0}$.

\par In case of a magnetic field line in form of a circle, $\vec{\kappa} = -1/R \, \vec{e}_r$ points towards the center. This also means that the magnetic tension force is always pointing to the the curve's center.

\par To summarize, the force due to the pressure gradient must be counteracted by the force due to the magnetic pressure, as well as the force due to the magnetic curvature \cite[75]{strothPlasmaphysikPhaenomeneGrundlagen2018}.

\section{Fusion Reactors}

\par As outlined in the beginning of the chapter, there is no confinement parallel to the field lines (excluding special magnetic gradient configurations employed in so-called mirror machines). Therefore, one bends the magnetic field lines to create a field line loop. One option is to bend them in form of a torus. 

\subsection{Tokamak Geometry}

\begin{figure}
  \center{\includegraphics[width=1.0\textwidth]{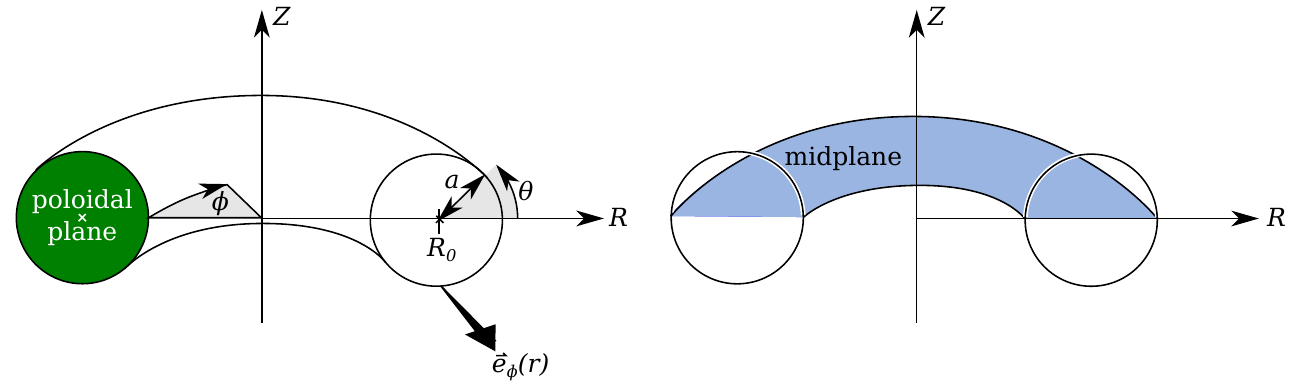}}
  \captionof{figure}{Reactor geometric labels}
  \label{fig:reactor_geometry}
\end{figure}

\par The main geometric parameters are the minor radius $a$ and the major radius $R_0$, in case the reactor is a perfect torus.

\par Poloidal refers to vectors lying in the poloidal plane (shown in green in figure \ref{fig:reactor_geometry}). Toroidal means parallel to the toroidal unit vector $\vec{e}_\phi(r)$. The term \enquote{axisymmetric} refers to symmetry around the $Z$ axis as shown in figure \ref{fig:reactor_geometry}.

\par Another geometric part is the midplane, which is obtained by slicing the torus into an upper and a lower half. It is marked in blue in figure \ref{fig:reactor_geometry}.

\subsection{Ohm's Law and the ExB Drift}
\label{ohms_law_ExB}

\par Ideal \acrshort{mhd} is called ideal, because it has no dissipative terms \cite[68]{militelloBoundaryPlasmaPhysics2022}. For example, no electrical resistivity is included. Inside a perfect conductor, the electric field is always zero: $\vec{E} = 0$ (ignoring acceleration and deceleration of charge carriers needed to adapt to electric field changes, as well as finite elementary charge). This is, because the charge carriers quickly move to a position that cancels the electric field. If the conductor is moved with velocity $\vec{v}$ inside a magnetic field, a Lorentz transformation has to be applied to this law, leading to $\vec{E}' = \vec{E} + \vec{v} \times \vec{B}$. Plugging $\vec{E}'$ into $\vec{E} = 0$ leads to ideal MHD's Ohm's law, which is given in equation \ref{eqn:mhd_ohms_law}.

\begin{equation}
\label{eqn:mhd_ohms_law}
\vec{E} = -\vec{v} \times \vec{B}
\end{equation}

\par In case of an induced electric field inside a perfect conductor, the charge carriers also
form a current that opposes the corresponding magnetic field change which in turn cancels the induced electric field, thus causing the electric field $\vec{E}$ to become zero. In a plasma with zero resistivity as in ideal \acrshort{mhd}, the particles can move freely in parallel direction. This means, that a perfectly conducting plasma counteracts any magnetic field changes. Instead, the magnetic field lines can only be bent, but not broken, by moving them together with the plasma, which is referred to as the frozen-in-field-line concept. A quantitative derivation of this is given in \cite[299-300]{freidbergPlasmaPhysicsFusion2007}.

\par By applying the vector identity $\vec{a} \times (\vec{b} \times \vec{c}) = \vec{b} (\vec{c} \cdot \vec{a}) - \vec{c} (\vec{a} \cdot \vec{b})$, one can take the cross product of equation \ref{eqn:mhd_ohms_law} with $\vec{B}$. This leads to $(\vec{v} \times \vec{B}) \times \vec{B} = -B^2 (\vec{v} - \vec{b} (\vec{b} \cdot \vec{v})) = -B^2 \vec{v}_\perp$. The resulting velocity is the so-called ExB velocity, better known as ExB drift:

\begin{equation}
\vec{v}_\mathrm{ExB} = \frac{\vec{E} \times \vec{B}}{B^2}
\end{equation}

\par It is noteworthy, that this drift is not dependent on the charges $q$ of the particles.

\par The ExB drift can also be derived more generally by resorting to single particles, instead of using \acrshort{mhd}. Here, the so-called guiding center is regarded, which is the center of the perpendicular gyration performed by a particle. Due to the gyration, the perpendicular component of any force $\vec{F}$ acting on the particle results in a drift of its guiding center \cite[19]{strothPlasmaphysikPhaenomeneGrundlagen2018}

\begin{equation}
\vec{v}_D = \frac{\vec{F} \times \vec{B}}{q B^2} \mathrm{.}
\end{equation}

\par Plugging in $\vec{F}_e = q E$, this leads to the single-particle velocity $\vec{v}_{D,\mathrm{center}} = \vec{E} \times \vec{B} / B^2$. The corresponding \acrshort{mhd} velocity is the mass-weighted average for ions and electrons and therefore has the same value. There are also other drifts not discussed here, like the $\nabla$B drift which results from non-uniform magnetic fields, and the curvature drift which results from curved magnetic fields. The force due to gravitation is so small compared to other forces, that it is neglected. 

\subsection{Diamagnetic Current}
\label{diamagnetic_current}

\par Another important phenomenon is the diamagnetic current. It can be calculated here by taking the cross product of the force balance (\ref{eqn:basic_mhd}) and $\vec{B}$. One can apply the same vector identity as for the ExB drift to get $(\vec{J} \times \vec{B}) \times \vec{B} = -B^2 \vec{J}_\perp$, which leads to the diamagnetic current 

\begin{equation}
\label{eqn:diamagnetic_current}
\vec{J}_\mathrm{dia} = -\frac{\vec{\nabla}p \times \vec{B}}{B^2} \mathrm{.}
\end{equation}

\par This equation can also be described as the \acrshort{mhd} force balance in perpendicular direction. It is called diamagnetic current, because it always opposes the initial magnetic field.

\begin{figure}
  \center{\includegraphics[width=0.15\textwidth]{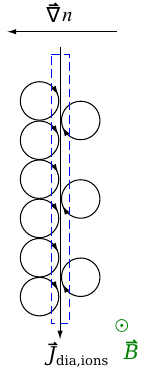}}
  \captionof{figure}{Ion part of the diamagnetic current due to the gyroscopic rotation around the magnetic field lines \cite[257]{freidbergPlasmaPhysicsFusion2007}.}
  \label{fig:magnetization_current}
\end{figure}

\par A more intuitive understanding can be gained with the help of figure \ref{fig:magnetization_current}. This shows the gyration of ions with decreasing density in a magnetic field. When only looking on the vertical surface element between the gyration centers marked by the dashed blue rectangle, there are more ions gyrating downwards than upwards, leading to the diamagnetic current of the ions. Since a temperature gradient does not cause less circles as for the density gradient, but smaller velocities, it will have a similar influence. To consider both temperature and density gradients, the pressure $p = n \, T$ is used in equation \ref{eqn:diamagnetic_current}. For electrons, a similar diamagnetic current occurs in the same direction. Both the ion and electron diamagnetic current then add up to form the diamagnetic current given by equation (\ref{eqn:diamagnetic_current}).

\begin{figure}
  \center{\includegraphics[width=0.5\textwidth]{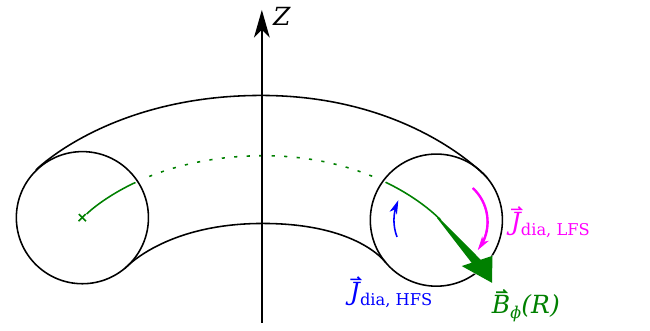}}
  \captionof{figure}{Diamagnetic currents $J_\mathrm{dia}$ in a reactor with purely toroidal magnetic field}
  \label{fig:diamagnetic_charge_buildup}
\end{figure}

\par On a torus with purely toroidal magnetic field, the diamagnetic current is purely poloidal. Also, the magnetic field near the torus center is bigger than at the outside of the torus due to the field line bending. Therefore, the section where $R < R_0$ is called the high-field side (\acrshort{hfs}), while the remaining space, where $R > R_0$, is called the low-field side (\acrshort{lfs}). An unwanted consequence of this is that the  \acrshort{hfs} dielectric currents are smaller than the \acrshort{lfs} dielectric currents, as indicated in figure \ref{fig:diamagnetic_charge_buildup}. This leads to a charge build-up. The electric field due to the charge difference creates an ExB drift that in turn leads to an expansion of the whole torus in $R$-direction, rendering the configuration unstable \cite[283-284]{freidbergPlasmaPhysicsFusion2007}. 

\par
\begin{figure}
  \center{\includegraphics[width=0.5\textwidth]{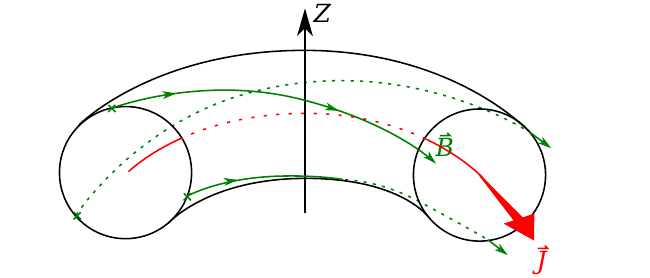}}
  \captionof{figure}{Twisted field line. All green lines correspond to a single field line that winds three times around the torus in toroidal direction.} 
  \label{fig:twisted_field_line}
\end{figure}

 This problem is solved by twisting the field lines, as indicated in figure \ref{fig:twisted_field_line}. Since particles can move freely along field lines, so-called Pfirsch-Schlüter currents form that build down the charge \cite[346]{strothPlasmaphysikPhaenomeneGrundlagen2018}. The shown field line has a safety factor $q$ of 3. This means, that it one poloidal turn, it makes three toroidal turns. The safety factor can be calculated by means of the following formula \cite[287-288]{freidbergPlasmaPhysicsFusion2007}:

\begin{equation}
\label{eqn:safety_factor}
q(r_0) = \frac{1}{2 \pi} \int_0^{2 \pi} \frac{r(\theta) B_\phi}{R B_\theta} \mathrm{d} \theta \mathrm{,}
\end{equation}

where $r(\theta)$ must be determined by following a magnetic field line along the torus, starting at ${r(\theta_0) = r_0}$.

\subsection{Tokamak and Stellarator Magnetic Field Configuration}
\label{tokamak_configuration}

\par There are two common approaches to achieve twisted field lines. The first one is called stellarator. In a stellarator, typically, a set of complicated non-planar coils is arranged to generate a twisted magnetic field.

\par The other approach is called tokamak. Here, a toroidal field is present initially due to poloidal coils wound around the torus, which are also called main field coils. Then, a current is created inside the plasma. This current generates a poloidal magnetic field, which results in twisted field lines when added to the toroidal magnetic field. The current is generated by means of a transformer coil situated in the center of the torus, which is called the central solenoid. This coil generates a field that increases over time such that a toroidal current is induced in the plasma, which is acting like a secondary winding in a transformer.

\par While the toroidal current is generated by induction in the simplest case, there is a neoclassical phenomenon\footnote{Neoclassical phenomena result from collisional behavior specific to a toroidal geometry. They can only be resolved by the single-particle picture, which shows that there exist so-called trapped particles with high perpendicular velocity compared to their parallel velocity, that cannot follow the twisted field lines but instead are reflected on their way towards the torus center, forming so-called banana orbits. Major results are much higher diffusion coefficients than for cylindrical geometries, as well as the bootstrap current. It must be noted that the diffusivities reconstructed from experimental fusion plasmas are often higher than predicted by neoclassical analysis due to micro-turbulences \cite[478-490]{freidbergPlasmaPhysicsFusion2007}.} called the bootstrap current, which flows parallel to the toroidal current and is self-generated by the plasma. With careful tailoring of the discharge design, a fully non-inductive toroidal current can be reached, of which a large fraction is contributed by the bootstrap current \cite{bockNoninductiveImprovedHmode2017}.

\par From here onwards, only plasmas confined in a tokamak will be regarded.

\subsection{Unstable Modes}

\par For stability analysis, it is convenient to Fourier decompose the displacement in toroidal direction and poloidal direction. In toroidal direction, the index $n$ is used to determine the so-called mode number, while in poloidal direction, the index $m$ is used:

\begin{equation}
\label{eqn:unstable_fourier}
f(x) = A \, \exp(2 \pi \, n \, i \, \phi) \exp(2 \pi \, m \, i \, \theta)
\end{equation}

\subsubsection{Ballooning Mode}

\par
\begin{figure}
  \center{\includegraphics[width=0.8\textwidth]{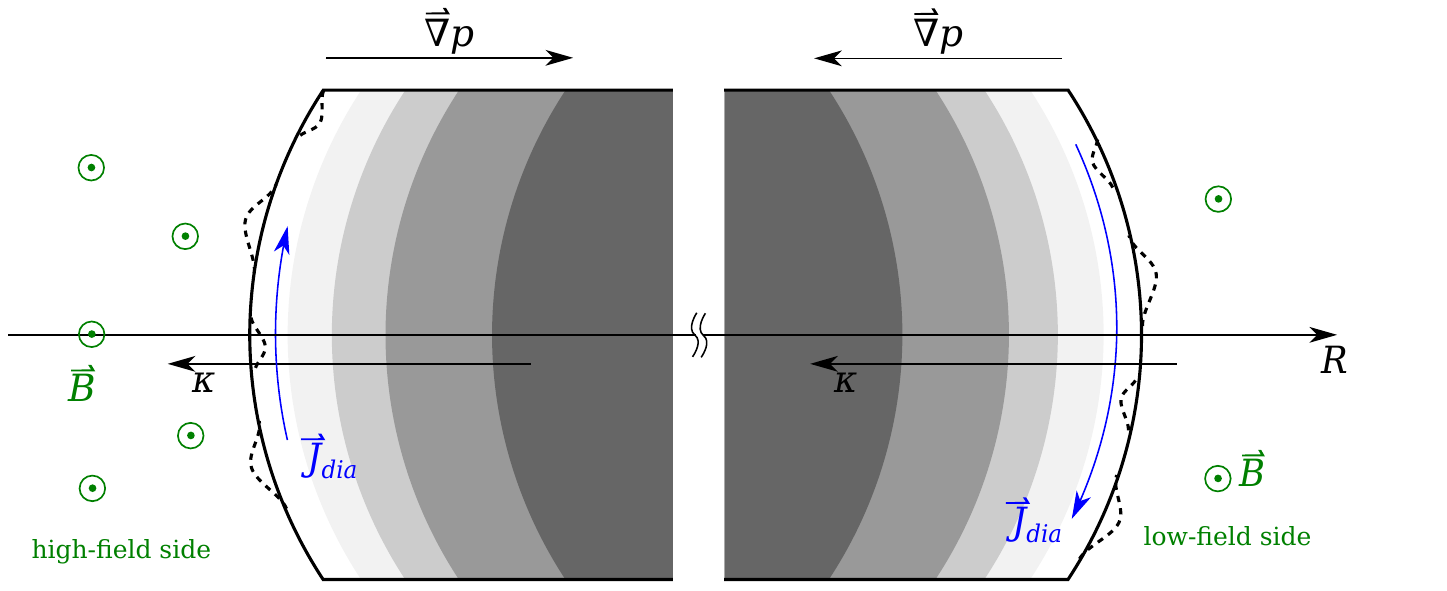}}
  \captionof{figure}{Ballooning mode start with dotted initial (small) perturbation. Note that the magnetic field $\vec{B}$ also exists inside the plasma, but was omitted for readability.}
  \label{fig:ballooning_mode1}
\end{figure}

 For stability analysis of any kind of mode, it is analyzed which initial perturbations will grow ($\rightarrow$ unstable mode), shrink back to the unperturbed case ($\rightarrow$ stable mode), or persist (e.g., oscillate). In figure \ref{fig:ballooning_mode1}, such an initial small perturbation is drawn dotted. It will be analyzed in the context of ballooning modes.

\begin{figure}
  \center{\includegraphics[width=0.8\textwidth]{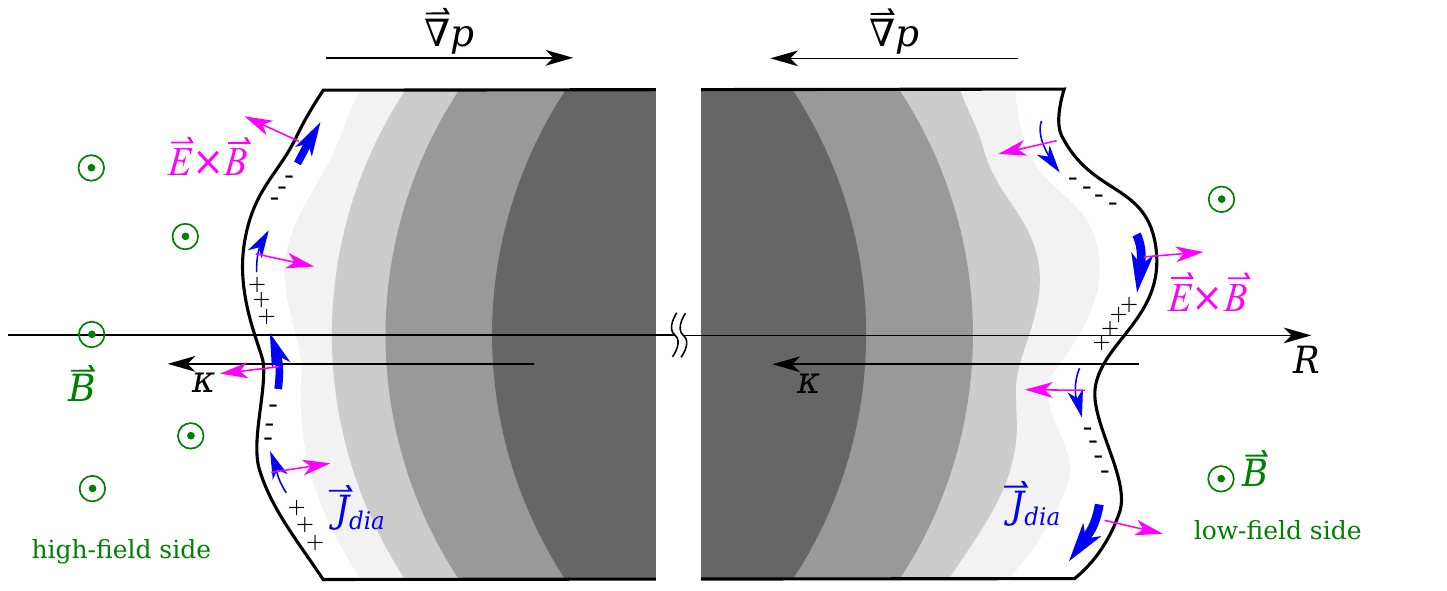}}
  \captionof{figure}{Ballooning mode evolution on the high-field side and low-field side. Thicker blue arrows represent higher diamagnetic current. Note that the magnetic field $\vec{B}$ also exists inside the plasma, but was omitted for readability.}
  \label{fig:ballooning_mode2}
\end{figure}

\par On the \acrshort{lfs}, this small perturbation enters a region where $B$ gets smaller when the plasma expands outwards. Due to that, the diamagnetic current $j_\mathrm{dia}$ increases, the farer the plasma expands outwards \cite[118]{strothPlasmaphysikPhaenomeneGrundlagen2018}\footnote{The ballooning mode is also accessible from the single-particle picture. There, the diamagnetic current is not present, but instead, the curvature drift $v_k = \mp \frac{v_\parallel^2 m}{q \, B} \frac{\vec{R} \times \vec{B}}{R_C^2 B}$ (upper sign for electrons, lower sign for ions) leads to a charge build-up equal to the one caused by the diamagnetic current changes.}. In contrast to that, when bulging inward, the plasma experiences higher $B$, leading to smaller $j_\mathrm{dia}$. The difference in diamagnetic current leads to a charge deposition between the bulges, as indicated in figure \ref{fig:ballooning_mode2}. Between the deposited charges, electric fields form. The resulting ExB drifts are in the bulging direction, nourishing further growth of the mode.

\par It should be noted that, although the magnetic field component of the diamagnetic current is $\propto 1/B \propto R$, and therefore increasing towards the outside as mentioned, there is also a pressure gradient component of $J_\mathrm{dia}$ that is $\propto \vec{\nabla}p$. One might expect that due to the expansion in the outwards direction, the pressure gradient is reduced, counteracting the increase of the diamagnetic current. Similarly, one would expect that in the inwards direction, the pressure gradient increases, counteracting the decrease of the diamagnetic current. But this expectation does not fully correspond to reality. Instead, due to the bulging, the mode increases only the surface, but not the volume, leading to smaller changes in the pressure gradient than expected.

\par
\begin{figure}
  \center{\includegraphics[width=0.55\textwidth]{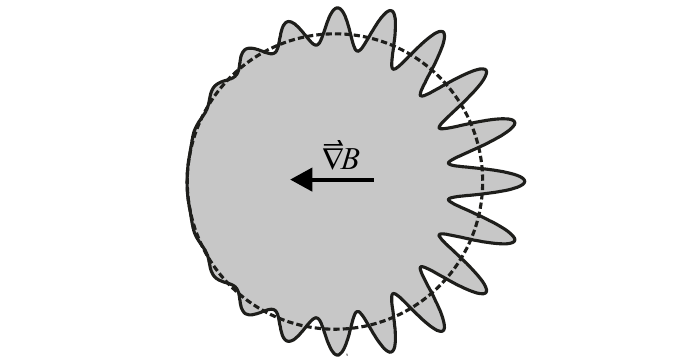}}
  \captionof{figure}{Ballooning mode typical appearance \cite[404]{strothPlasmaphysikPhaenomeneGrundlagen2018}}
  \label{fig:stroth_ballooning_mode}
\end{figure}

 On the \acrshort{hfs}, the diamagnetic current is higher towards the inside of the plasma. This leads to an ExB drift that counteracts the mode instead of increasing it as on the \acrshort{lfs}. The resulting appearance due to growth at the \acrshort{lfs} and suppression at the \acrshort{hfs} is sketched in figure \ref{fig:stroth_ballooning_mode}.

\par The difference between the \acrshort{hfs} and \acrshort{lfs} behavior can be summarized by regarding the direction of the curvature vector $\kappa$, which is also shown in figures \ref{fig:ballooning_mode1} and \ref{fig:ballooning_mode2}. If $\kappa$ is parallel to $\vec{\nabla}p$ (as for the \acrshort{lfs}), this is called unfavorable curvature, because it promotes ballooning modes. When they are antiparallel, it is called favorable curvature, because it suppresses ballooning modes.

\par An important property of the ballooning mode is a typically small toroidal mode number $n$. Po\-lo\-i\-dal\-ly, the mode can be decomposed into many modes around a typically large $m$, with constructive interference at the \acrshort{lfs} and destructive interference at the \acrshort{hfs}. Note that for ballooning modes, the poloidal part of the Fourier decomposition as written in equation (\ref{eqn:unstable_fourier}) is modified slightly using the so-called ballooning transformation as described in \cite[20]{breuLinearNonlinearAnalysis2011}.

\subsubsection{Kink Mode}

\par For the ballooning mode, $j_\perp$ and $B_\parallel$ have been regarded. In contrast to that, $j_\parallel$ and $B_\perp$ are the drivers for kink modes. For kink modes to become unstable, a high $j_\parallel$ must be present in the plasma. This is necessary because $B_\perp$ must be mostly self-generated (as opposed to the external generation by coils as done in some other reactor configurations, e.g., stellarators) to excite the mode.

\begin{figure}
  \center{\includegraphics[width=0.4\textwidth]{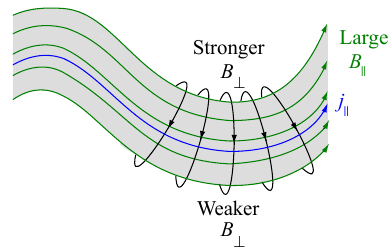}}
  \captionof{figure}{Kink instability principle (adapted from \cite[305]{freidbergPlasmaPhysicsFusion2007})}
  \label{fig:freidberg_kink_instability}
\end{figure}

\par Figure \ref{fig:freidberg_kink_instability} shows the basic principle for an $n=1, m=1$ kink mode. An initial perturbation is present, that moves the plasma off-center perpendicularly. The parallel current $j_\parallel$ is moved by this, too. Therefore, the perpendicular magnetic field $B_\perp$ generated by $j_\parallel$ is moved as well.

\par Due to the sinusoidal perturbation, $B_\perp$ is rotated together with the plasma, causing stronger $B_\perp$ fields in the newly created curve centers. Therefore, $\vec{j}_\parallel \times \vec{B}_\perp$ is higher towards the newly formed curve center than towards the direction of perturbative movement, causing a net force that moves the plasma further.

\begin{figure}
  \center{\includegraphics[width=0.4\textwidth]{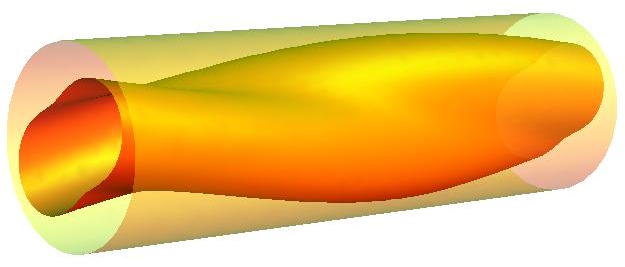}}
  \captionof{figure}{$n = 1,\ m = 2$ kink instability \cite{cowleyLectureKink}}
  \label{fig:cowley_m2_n1_kink}
\end{figure}

\par As described in section \ref{tokamak_configuration}, a large fraction of $\vec{j}_\parallel$ is due to the bootstrap current. This current is large at high pressure gradients, which applies especially to the edge. This is the reason for why the bootstrap current is a driving factor for kink modes \cite[23]{breuLinearNonlinearAnalysis2011}.

\par Figure \ref{fig:cowley_m2_n1_kink} visualizes a kink instability for $m = 2$. A property of the kink mode is its typically small toroidal mode number $n$.

\subsubsection{Peeling-Ballooning Mode}

\par
\begin{figure}
  \center{\includegraphics[width=0.7\textwidth]{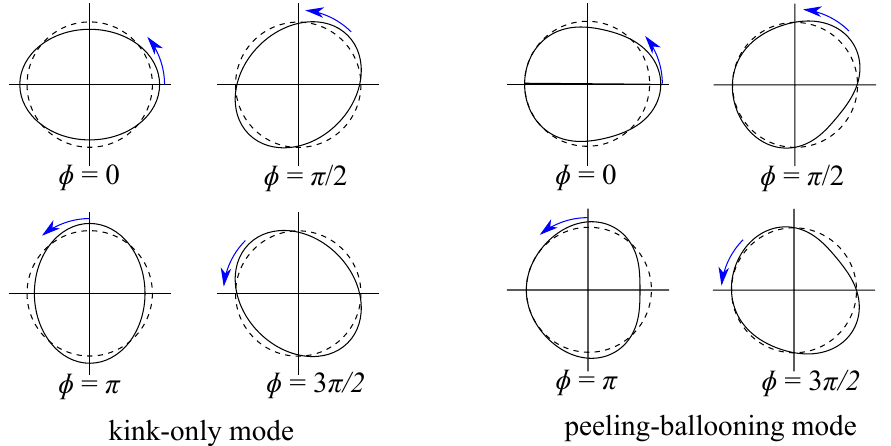}}
  \captionof{figure}{$n = 1,\  m = 2$ peeling-ballooning mode, followed along the torus (adapted from \cite[396]{freidbergPlasmaPhysicsFusion2007}). Blue arrows indicate the kinking direction.}
  \label{fig:peeling_ballooning_n1_m2}
\end{figure}

 In a tokamak, peeling and ballooning mode combine, leading to a kinking-like profile that bulges more on the outside of the torus than at the inside, as shown in figure \ref{fig:peeling_ballooning_n1_m2}.

\subsubsection{Edge-Localized Modes (\acrshort{elm}s)}

\par Peeling-Ballooning modes driven by the large pressure gradients and the large current densities in the pedestal region at the boundary of the confined plasma region can non-linearly develop into edge-localized modes (\acrshort{elm}s). Thus, for H-mode plasmas, \acrshort{elm}s can occur repetitively. One typically distinguishes between the \acrshort{elm} crash, during which an ELM grows, until it leads to the ejection of material, and the inter-\acrshort{elm} phase, which takes place between consecutive \acrshort{elm} crashes. There are multiple types of \acrshort{elm}s, of which the potentially most destructive one is know as the type-I \acrshort{elm}. Its energy loss per occurrence is largest among the \acrshort{elm} types, ranging from 5 to 15\% of the total energy of the confined plasma before the \acrshort{elm}, which is the reason for why they are sometimes called large \acrshort{elm}s. Due to the very large amount of energy released by type I \acrshort{elm}s, they can pose a risk to the reactor structures. Therefore, they require mitigation and avoidance schemes for upcoming machines like ITER \cite{eichELMDivertorPeak2017}.

\par In addition to the high peak loads on the divertor plates, they also have the drawback that they can limit the highest reachable pedestal top pressures, temperatures and densities, which limits the confinement in the core as well. This is because \acrshort{elm}s typically occur when the pressure gradient crosses a certain threshold (dependent on other parameters like triangularity of the cross-section), flushing parts of the pedestal \cite[1]{snyderEdgeLocalizedModes2002}. This is the motivation for why this thesis focuses on type-I \acrshort{elm}s (especially on their onset).

\par Ref. \cite{snyderEdgeLocalizedModes2002} shows, that the conditions for pressure gradient and toroidal pedestal current, under which large \acrshort{elm}s occur, match the predicted stability boundaries of high-$n$ pure ballooning modes, intermediate-$n$ peeling-ballooning modes and low-$n$ peeling modes (with the least stable modes dictating the boundary). This is one of the indications that these \acrshort{elm}s are indeed initiated by peeling/ballooning modes, which in turn can be accessed by \acrshort{mhd} theory. While for the analysis of the onset of an \acrshort{elm}, linear \acrshort{mhd} simulations\footnote{In linear simulations, perturbation theory is used: The plasma quantities like magnetic field, temperature, density (therefore pressure) and others are written like $\vec{B} = \vec{B}_0 + \vec{B}_1$, $T = T_0 + T_1$ and so on, with the equilibrium values $\vec{B}_0$, $T_0$, ... and the perturbations $\vec{B}_1$, $T_1$ etc. This assumes small perturbations compared to the equilibrium. In contrast to that, non-linear simulations do not use perturbation theory, and thus do not require the modes to be small compared to the equilibrium.} might be sufficient, the overall \acrshort{elm} dynamics like expelled energy and repetition frequency are highly non-linear and therefore require non-linear codes for simulation \cite[84]{zohmMagnetohydrodynamicStabilityTokamaks}\cite{catheyNonlinearMHDSimulations2021}.

\par This thesis will for the major part analyze the mostly linear ballooning and peeling modes growth that typically precedes type-I \acrshort{elm}s using a non-linear code, thus allowing to continue the mode evolution up to the actual non-linear occurrence of the \acrshort{elm} at the end of the thesis.

\section{Effects of Different Isotopes}
\label{different_isotopes_fundamentals}

\par There exist multiple larger-scale tokamak experiments, among them \acrshort{aug}, \acrshort{jet}, DIII-D, KSTAR, EAST and JT-60SA. Out of these, only \acrshort{jet} has ever run D-T or T-T campaigns. Reasons are, among others, that Tritium is very costly due to its scarcity and that it requires increased radiation protection measures.  

\par One such D-T campaign of \acrshort{jet} is described in \cite{frassinettiEffectIsotopeMass2023}. Here, reduced particle transport for higher ion mass $A_\mathrm{eff}$ is believed to be the cause for a measured increase of the pedestal height of the particle density. Unfortunately, the normalized pressure $\beta_N$ was experimentally found to be dependent on the particle density, but also was theoretically predicted to have a high influence on type-I \acrshort{elm}s as shown in \cite{snyderEdgeLocalizedModes2002}. Therefore, to keep $\beta_N$ constant, the heating power had to be reduced in the experiment. This, together with measurement uncertainties, made it impossible to check for differences in heat diffusivity $\chi$. For comparative experimental studies, these kind of couplings make it hard to understand how and why the \acrshort{elm} behavior is changing from deuterium to tritium. But the campaign shows, that T has at least not a higher heat transport than D, and that it is probable (but there is lack of sufficient data) that the heat transport is less for T than for D.

\par When comparing protium with deuterium, much more results are available, because the implementation issues associated with tritium do no longer apply. Ref. \cite{horvathIsotopeDependenceType2021} shows, that the experimentally measured pedestal profile differences in \acrshort{jet} can be mapped to a decrease of the heat and particle transport for increasing $A_\mathrm{eff}$, which agrees with the results of the D-T campaign mentioned before. The findings of \cite[3-4]{viezzerIonHeatTransport2018} fit to this, where it is shown that the measured heat diffusivities in \acrshort{aug} are also lower for deuterium, compared to protium. As will be explained later in more detail, this thesis will not incorporate these transport effects related to small-scale turbulence, which are not part of the \acrshort{mhd} model used here.

Ref. \cite[13]{horvathIsotopeDependenceType2021} also mentions that the growth rates from ideal \acrshort{mhd} $\gamma_\mathrm{MHD}$ scale with $1/\sqrt{A_\mathrm{eff}}$\footnote{While not explained in the paper, the reason is that when analyzing stabilities by calculating the potential energy, the formula $\gamma \geq \sqrt{\delta W}{K}$ gives a lower bound of the growth rate, with $K \propto \rho$ and $\delta W$ being independent from $\rho$ \cite[353]{freidbergIdealMHD2014}, which also serves as \enquote{reasonable estimate} by replacing $\geq$ with $\approx$ \cite[355]{freidbergIdealMHD2014}. The factor of $\rho$ in $K$ does arrive from the left side of ideal MHD's momentum equation.}. Because ideal \acrshort{mhd} does not incorporate the ion diamagnetic velocity, the condition for a mode to be stable is not $\gamma_\mathrm{MHD} > 0$ but $\gamma_\mathrm{MHD}/2 > \omega_{\star i}$, with the ion diamagnetic frequency $\omega_{\star i} = m/r \cdot T_i / (e_i B_0) \cdot d \ln p_i / dr$ with the poloidal mode number $m$ \cite[7]{snyderEdgeLocalizedModes2002}\cite[4296]{huijsmansDiamagneticStabilization2001}. Since the ion diamagnetic frequency is independent from the ion mass, but $\gamma_\mathrm{MHD}$ is not, this leads to increased stabilization, the higher the ion mass.

\par Section 10 of \cite{frassinettiEffectIsotopeMass2023} also summarizes that the increase of the particle density pedestal height (which is proportional to the fraction of separatrix and pedestal electron density $n_e^\mathrm{sep}/n_e^\mathrm{ped}$) leads to an increase in pedestal turbulent transport that reduces the temperature gradient and makes resistive effects non-negligible, citing \cite{frassinettiPedestalStructureStability2021} and \cite{nystromResistivityStability2022}. This has a non-negligible influence on \acrshort{elm} stability, especially at high power and high gas rate\footnote{the gas rate refers to the amount of gas injected into the plasma as a particle source}.

\par Not all profiles are directly impacted by the ion mass. For example, changes of the radial electric field $E_r$ are seen to be not dependent on it, but instead are correlated to the ion diamagnetic term $\nabla p / (n e)$ \cite[3]{viezzerParameterDependenceRadial2014}\cite[7]{schneiderOverviewIsotopeEffects2021}, which then in turn might be affected by the ion mass.

\par Finally, ref. \cite[13-14]{frassinettiEffectIsotopeMass2023} also shows that for increasing ion mass, the energy lost during an \acrshort{elm} increases as well. Experimentally, this is largely due to an increase of the particle density pedestal losses from around 25\% for deuterium to around 40\% for tritium in the shots which were analyzed.

\clearemptydoublepage
\chapter{Approach}
\label{approach}
\section{Fusion Plasma Modeling and Simulation}

\par To simulate fusion plasmas, one can use either a kinetic-based, or a fluid dynamics-based approach. The former treats each particle (each electron, ion and neutral particle) separately, to solve directly for the six-dimensional distribution function introduced in section \ref{force_balance_of_plasma}. Reducing the gyration motion of charged particles ultimately leads to a gyro-kinetic distribution function. It is very costly, but some groups do this (sometimes with reduced spatial domains instead of the entire plasma), using codes like GENE \cite{jenkoElectronTemperatureGradient2000}.  

\par One approach based on fluid dynamics is magnetohydrodynamics (\acrshort{mhd}),
as introduced in section \ref{force_balance_of_plasma}. This approach is computationally less expensive, but leaves out effects that fundamentally originate from particles, such as finite Larmor radius effects resulting from the gyration of particles around the field lines. Still, the remaining equation system is not easy to solve, requiring further simplification. 

\par The method used here and implemented by the JOREK code \cite{hoelzlJOREKNonlinearExtended2021} is visco-resistive \acrshort{mhd}, with certain extensions to extend its region of validity, and will be further described in the following.

\subsection{Non-linear Extended MHD Using JOREK}
\label{approach_jorek}

\par While JOREK is comprised of different models to study distinct physical processes, the model used here can be described as reduced \acrshort{mhd} with diamagnetic two-fluid extensions\footnote{The model is referred to as model303 in the code}.

\par While the geometrical variables like $R$ and $Z$, as well as the magnetic field $B$ and the poloidal magnetic flux $\psi$ are kept in SI units, most variables are normalized. Section \ref{normalization} will go into more detail about the normalization. From now on, only the normalized quantities will be used, while values in SI units will be decorated with an \enquote{SI} subscript. Also, the effective atomic mass number $A_\mathrm{eff}$ is used as the normalized ion mass, with ${m_{i, SI} = A_\mathrm{eff} \cdot m_\mathrm{proton}}$, where $m_\mathrm{proton} = 1.673 \cdot 10^{-27} \, \mathrm{kg}$ is the proton mass.

\par Reduced means here, that JOREK uses an ansatz where the toroidal part of the magnetic field is kept constant in time \cite[8]{hoelzlJOREKNonlinearExtended2021}\footnote{Poloidal currents can still exist, though, but the toroidal magnetic field created by them is neglected \cite[21]{catheyNonlinearMHDSimulations2021}.}. This eliminates fast magnetosonic waves from the system, which makes it easier to solve it numerically \cite[1334]{franckEnergyConservationNumerical2015}. While some instabilities can be affected by keeping $B_\mathrm{tor}$ constant, the peeling and ballooning modes which are relevant for \acrshort{elm}s are not influenced by it \cite{pamelaExtendedFullMHDSimulation2020}\footnote{Also note that instabilities like the ballooning and peeling modes described in section \ref{plasma_physics} are typically embodied by changes of the poloidal magnetic field rather than changes of the toroidal magnetic field.}. This ansatz is applied by setting the magnetic vector potential $\vec{A} = \psi \frac{1}{R} \vec{e}_\phi$ with $\vec{B} = B_\phi(R) \vec{e}_\phi + \vec{\nabla} \times \vec{A}(r, t)$.

\par Based on the constant toroidal magnetic field, one can formulate an additional ansatz for the velocity, which allows for further simplifications. Using the velocity stream function $u = \Phi / F_0$, with $F_0 = R_0 B_{\phi, 0}$, and $\Phi$ being the electrostatic potential, one can state \cite[3]{pamelaRecentProgressQuantitative2017}

\begin{equation}
\label{eqn:velocity_without_diamagnetic}
\vec{v} = -R \vec{\nabla}u \times \vec{e}_\phi + v_\parallel \vec{B} \mathrm{,}
\end{equation}

neglecting the poloidal resistivity\footnote{Originally, Ohm's law is $\vec{E} + \vec{v} \times \vec{B} = \frac{1}{\rho}(\vec{J} \times \vec{B} - \vec{\nabla}{p}) + \eta \vec{J}$ \cite[250]{freidbergPlasmaPhysicsFusion2007}. Not only is the resistivity term $\eta \vec{J}$ neglected, but also the hall term $\frac{1}{\rho}\vec{J} \times \vec{B}$, as well as the diamagnetic term $\frac{1}{\rho}\vec{\nabla}{p}$. The latter one is added again back to the equations later on in this thesis. It should be noted that the hall term can be neglected because the velocity ansatz represents gyro-viscous cancellation, which expresses Ohm's law in terms of the ExB-drift velocity $\vec{v}_E = \vec{E} \times \vec{B} / B^2$ instead of the total velocity $\vec{v}$. The remaining $\vec{J} \times \vec{B}$ term in Ohm's law then becomes very small \cite[302]{schnackLecturesMHD2009}.}. The full derivation can be found in \cite[21-22]{catheyNonlinearMHDSimulations2021}.

\par In JOREK's reduced \acrshort{mhd} variant, there is implemented a system of equations that is solved for the poloidal magnetic flux $\psi$, mass density $\rho$, temperature $T$ (with $p = \rho T$ in normalized units), the, just introduced, velocity stream function $u$, and the fluid velocity along magnetic field lines $v_\parallel$. The system is explicitly solved for the toroidal vorticity $\omega_\phi$ and toroidal current density $j_\phi$, too. These are auxiliary, because $\omega_\phi$ is essentially a derivative of $\Phi$ ($\omega_\phi = \vec{\nabla} \cdot \vec{\nabla}_\mathrm{pol} u$), and $j_\phi$ is essentially a derivative of $\psi$ ($j_\phi = R^2 \vec{\nabla} \cdot (1 / R^2 \vec{\nabla}_\mathrm{pol} \psi)$). They are kept for numerical reasons \cite[10]{hoelzlJOREKNonlinearExtended2021}: while directly using higher-order derivatives in the implementation can introduce noise, this can be avoided by plugging in $j$ and $\omega$ instead (or first-order derivatives thereof), and adding their definitions to the set of equations which are solved by the code.

\par For ideal \acrshort{mhd}, it was assumed, that the fluid was a perfect electrical conductor, and that it had neither viscosity nor thermal conductivity \cite[49]{schnackLecturesMHD2009}. Extended \acrshort{mhd} does not require the stated assumptions anymore. Essentially, this results in extra terms that are added to the ideal \acrshort{mhd} equations \cite[285-296]{schnackLecturesMHD2009}. For example, starting from the momentum equation (\ref{eqn:basic_mhd}), JOREK adds a viscosity term $\mu \Delta \vec{v}$ which was previously neglected \cite[3]{huysmansNonlinearMHDSimulations2009}, resulting in equation (\ref{eqn:jorek_momentum}). Here, $\mu$ is the visosity. Additionally, a source term $\vec{S}_V$ is present to account for density changes. This is described in more detail in \cite[8]{hoelzlJOREKNonlinearExtended2021}. The term $\rho \vec{v} \cdot \nabla \vec{v}$ is present because equation (\ref{eqn:jorek_momentum}) is formulated in the rest frame (as opposed to the \acrshort{mhd} frame that is moving with the fluid element).

\begin{equation}
\label{eqn:jorek_momentum}
\rho \frac{\partial \vec{v}}{\partial t} = -\rho \vec{v} \cdot \nabla \vec{v} + \vec{J} \times \vec{B} - \vec{\nabla} p + \mu \Delta \vec{v} + \vec{S}_V
\end{equation}

\par The density equation of the JOREK code,

\begin{equation}
\label{eqn:jorek_density}
\frac{\partial \rho}{\partial t} = -\vec{\nabla} \cdot (\rho \vec{v}) + \vec{\nabla} \cdot (\underline{D} \vec{\nabla} \rho) + S_\rho \mathrm{,}
\end{equation}

describes the change in density due to mass movement ($\rho \vec{v}$) and particle diffusion ($\underline{D} \vec{\nabla} \rho$). Since turbulent and neoclassical transport is not resolved by the simulation, this kind of diffusion is applied by setting a matching diffusion coefficient. The density equation also includes a particle source term, $S_\rho$, that describes the creation of new plasma particles. In experimental conditions, these arise from ionisation, for example \cite[24]{catheyNonlinearMHDSimulations2021}. The heat equation, which can be found in \cite{hoelzlJOREKNonlinearExtended2021}, includes a heat diffusivity term and heat source term for the same reasons.

\par JOREK also includes a resistivity term which results in an additional component of Ohm's law

\begin{equation}
\vec{E}_\mathrm{resis} = \eta \vec{J} \mathrm{,}
\end{equation}

where $\eta$ is the resistivity. The plasma can then not only bend, but also break field lines, so that the frozen-in-field-line concept no longer applies (see section \ref{ohms_law_ExB}). This gives rise to more instabilities, as well as modifying the growth rates and non-linear dynamics of ideal instabilities \cite[41]{breuLinearNonlinearAnalysis2011}\cite[124]{zohmMagnetohydrodynamicStabilityTokamaks}.

\par Another important component is the diamagnetic velocity, which is related to the diamagnetic current described in section \ref{diamagnetic_current}. This is expressed as

\begin{equation}
\vec{E}_\mathrm{diamag} = F_0 \frac{\delta^\star}{\rho} (\vec{\nabla}_\perp p_i) \mathrm{,}
\end{equation}

where $\delta^\star_{SI} = (\Omega_{c i} R_0)^{-1}$ with the ion gyrofrequency $\Omega_{ci} = e B_0 / m_{i, SI}$ \cite[3]{pamelaRecentProgressQuantitative2017}. In normalized units, $\delta^\star$ can be written as $A_\mathrm{eff} \, m_p / (e \, F_0 \sqrt{\mu_0 \rho_0})$ by multiplying it with a factor of $1 / \sqrt{\mu_0 \rho_0}$ due to normalization. The diamagnetic current also leads to an additional term for the total velocity $\vec{v}$ in equation (\ref{eqn:velocity_without_diamagnetic}), which becomes

\begin{equation}
\vec{v} = -R \vec{\nabla}u \times \vec{e}_\phi + v_\parallel \vec{B} - \frac{\delta^\star R}{\rho} \vec{\nabla}p_i \times \vec{e}_\phi \mathrm{.}
\end{equation}

\par Due to single-fluidity, the ion pressure $p_i$ is not available and is instead assumed to be half the total pressure $p$ \cite[6]{hoelzlJOREKNonlinearExtended2021}, leading to

\begin{equation}
\vec{E}_\mathrm{diamag} = F_0 \frac{\tau_{IC}}{\rho} (\vec{\nabla}_\perp p) \mathrm{,}
\end{equation}  

with the constant $\tau_{IC} = \delta^\star / 2 = A_\mathrm{eff} \, m_p / (2 e \, F_0 \sqrt{\mu_0 \rho_0})$.

\par In JOREK, the bootstrap current is implemented using the Sauter model \cite{sauterNeoclassicalConductivityBootstrap1999}.

\par In this thesis, an ideally conducting wall is assumed (all perturbations of $\psi$ and $j$ at the computational boundary set to zero). At the divertor, the field lines cross the boundary. Therefore, this condition is not applied there. Instead, the parallel component of the velocity is set to the local speed of sound\footnote{To calculate the ion speed of sound for the boundary condition, JOREK uses the formula $c_{i, SI} = \sqrt{\frac{\gamma \, T_{SI}}{m_{i, SI}}}$.
}, and the particles (density) and heat (temperature) are allowed to flow freely as described in \cite[3]{huysmansNonlinearMHDSimulations2009}. Note that JOREK is capable of simulating a resistive wall by coupling to the STARWALL or CARIDDI code, but an ideal wall boundary condition was used here instead for simplicity.

\par Finally, it should be noted that fusion effects, such as the generation of fast Helium ions, are not incorporated into the simulation.

\subsection{Further Notation}

\par The normalized poloidal flux is defined with the purpose of providing a radial coordinate as $\psi_N = (\psi - \psi_\mathrm{axis}) / (\psi_\mathrm{bnd} - \psi_\mathrm{axis})$, so that it becomes 1 at the separatrix and 0 at the core. This implies that due to small changes in the poloidal flux, profiles dependent on the poloidal flux (like heating and particle sources as explained in the next chapter) will only approximately stay constant when plotted over the radius $r$. Note that this does not apply to the toroidal magnetic field $B_\parallel$, which stays constant over time with respect to $R$.

\clearemptydoublepage
\chapter{Results}
\label{results}
\par In this chapter, JOREK simulations comparing only the effect of changing the ion mass are presented and contrasted. These comparisons cannot be achieved in the experiment, because the underlying transport due to microturbulences and neoclassical effects changes the circumstances and makes it impossible to have similar pedestals for different ion masses. The overall structure of the chapter is the following: Section \ref{normalization} explains the setup, its differences between the ion masses, as well as the evolution of the profiles during the simulation when not including possibly unstable harmonics. Section \ref{particle_source_modification} details out why the particle source had to be adapted for each ion mass. Section \ref{single_n_simulation} then moves on with simulations of a single toroidal mode for the D case only. This is followed by section \ref{single_n_growth_rate_comparison}, which compares the modes between D, D-T and T. In section \ref{momentum_equation_explanation}, this difference is investigated further by applying artificial modifications on the \acrshort{mhd} model. Finally, in section \ref{multi_n_simulation}, a configuration with many toroidal harmonics is used to simulate and compare an \acrshort{elm} crash between the isotopes.

\section{Simulation Setup for Different Isotopes}
\label{normalization}

\par The overall goal of the comparative simulations is, as described, to nail down the effects resulting from \acrshort{mhd} due to different ion masses. It was chosen to set up a simulation for $A_\mathrm{eff} = 2.0$, corresponding to D, $A_\mathrm{eff} = 2.5$, corresponding to a mixture of 50\% D and 50\% T, and $A_\mathrm{eff} = 3.0$, corresponding to a pure T plasma. In ASDEX Upgrade, tritium cannot be used experimentally in practice due to radiation protection, but this work can anyway only aim for a qualitative study, not yet for quantitative experiment comparisons.

\begin{figure}
  \center{\includegraphics{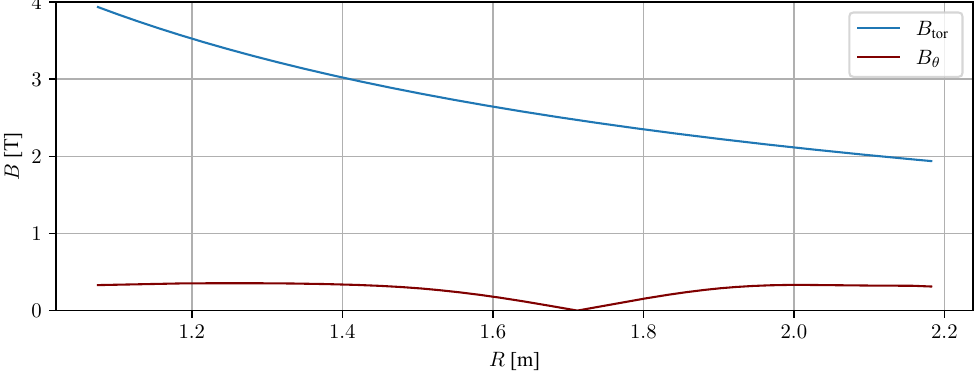}}
  \captionof{figure}{Constant toroidal magnetic field, and initial profile of the poloidal magnetic field (absolute value). The latter one evolves slightly over time due to toroidal current build-up.}
  \label{fig:B_tor}
\end{figure}

\par The geometry of the simulated reactor is based on the (axisymmetric) tokamak experiment \acrshort{aug}. The initial profiles were extracted in \cite{catheyNonlinearMHDSimulations2021} from \acrshort{aug} discharge \#33616 at roughly 7\,s using the CLISTE code \cite{patrickCLISTEInterpretiveEquilibrium1999}.

\par
\begin{figure}
  \center{\includegraphics[width=0.5\textwidth]{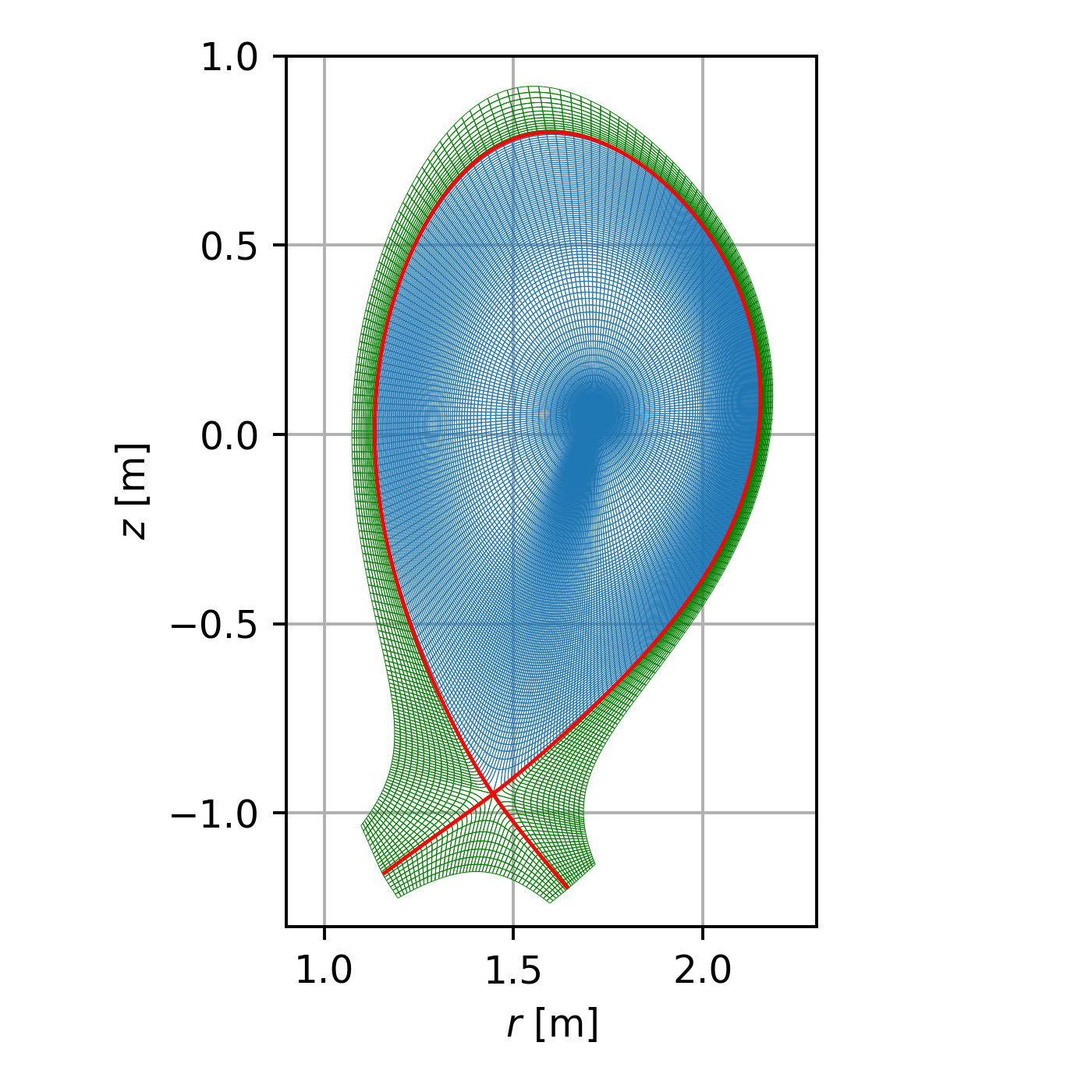}}
  \captionof{figure}{Simulation grid. Confined plasma indicated in blue, separatrix indicated in red, and scrape-off layer indicated in green.}
  \label{fig:grid}
\end{figure}

\begin{figure}
  \center{\includegraphics{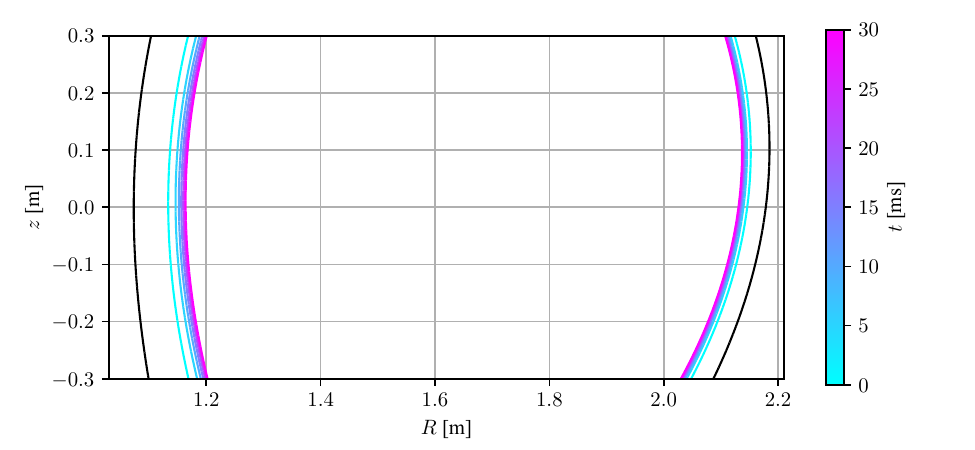}}
  \captionof{figure}{Inwards movement of the separatrix over time. The computational boundary, which is represented by a perfectly conducting wall, is indicated in black.}
  \label{fig:separatrix_evolution}
\end{figure}

 The stationary toroidal magnetic field generated by the main field coils is plotted in figure \ref{fig:B_tor}. The initial profile of the poloidal magnetic field is also shown, which evolves over time. For initialization, JOREK calculates the Grad-Shafranov equilibrium that is based on ideal MHD, using the density, temperature and current\footnote{For the current, the actual input is $F\,\mathrm{d}F/\mathrm{d}\psi$, where $F$ is basically the poloidal current density.} profiles from \acrshort{aug}. The initial grid is a polar grid, but the grid is turned to a flux-surface aligned grid after finding the solution to the Grad-Shafranov equation (namely, the $R,Z$ map of the poloidal magnetic flux). This can be visualized in figure \ref{fig:grid}. There, the separatrix is drawn in red, separating the plasma (blue) from the outer volume (green). This grid is kept constant for the rest of the simulation. This means, that it slightly looses alignment with the magnetic flux surfaces, because the pressure gradient changes in the course of the simulation and therefore, the equilibrium is adapted. Figure \ref{fig:separatrix_evolution} illustrates this for the separatrix, which represents the flux surface (for which $\psi_N = 1$), that deviates from the original positions over time.  

\par To set up comparable simulations for different isotopes, the initial simulation for each isotope is configured such that the SI values of particle density $n$, velocity, pressure, temperature, electrical field, current and most input variables like diffusion coefficients and source coefficients are the same. As a consequence of using same particle densities, the mass density $\rho$ varies between cases.

\par Table \ref{tab:parameters_si} lists the used parameters in SI units, as well as the formulas from \cite{hoelzlJOREKNonlinearExtended2021} to calculate the values in normalized units for each ion mass case.

\begin{table}
\centering
\caption{Simulation parameters and their conversion to normalized units}
\begin{tabularx}{\textwidth}{p{4.3cm}|l|l|p{5.4cm}}
Parameter & Symbol & SI value & Conversion to normalized units \\ \hline
central resistivity & $\eta_{0, SI}$ & $6.649 \cdot 10^{-8} \Omega \mathrm{m}$ & $\eta_{0, SI} \cdot \sqrt{\rho_0 / \mu_0}$ \\ \hline
central perpendicular viscosity & $\mu_{\perp, 0, SI}$ & $1.0377 \cdot 10^{-8} \mathrm{kg} \mathrm{m}^{-1} \mathrm{s}^{-1}$ & $\mu_{\perp, 0, SI} \cdot \sqrt{\rho_0 / \mu_0}$ \\ \hline
parallel viscosity & $\mu_{\parallel, SI}$ & $1.0377 \cdot 10^{-7} \mathrm{kg} \mathrm{m}^{-1} \mathrm{s}^{-1}$ & $\mu_{\parallel, SI} \cdot \sqrt{\rho_0 / \mu_0}$ \\ \hline
perpendicular particle \newline diffusivity & $D_{\perp, SI}$ & see figure \ref{fig:D_perp} & $D_{\perp, SI} \cdot \sqrt{\mu_0 \rho_0}$ \\ \hline
parallel particle \newline diffusivity & $D_{\parallel, SI}$ & 0 & 0 \\ \hline
perpendicular heat \newline diffusivity & $\kappa_{\perp, SI}$ & see figure \ref{fig:kappa_perp} & $\kappa_{\perp, SI} \cdot \sqrt{\mu_0 / \rho_0} \cdot (\gamma - 1) \cdot m_{i, SI}$ \\ \hline
central parallel heat \newline diffusivity & $\kappa_{\parallel, 0, SI}$ & $8.092 \cdot 10^{30} \mathrm{m}^{-1} \mathrm{s}^{-1}$ & $\kappa_{\parallel, 0, SI} \cdot \sqrt{\mu_0 / \rho_0} \cdot (\gamma - 1) \cdot m_{i, SI}$ \\ \hline
heat source & $S_{E, SI}$ & see figure \ref{fig:heat_source} & $S_{T, SI} (\gamma - 1) \mu_0 \sqrt{\mu_0 \rho_0}$ \\ \hline
particle source & $S_{\rho, SI}$ & see figure \ref{fig:particle_source} & $S_{\rho, SI} \cdot \sqrt{\mu_0 / \rho_0}$ 
\end{tabularx}
\label{tab:parameters_si}
\end{table}

\par The normalization density required for the conversion, $\rho_0 = A_\mathrm{eff} \cdot n_0 \cdot m_\mathrm{proton}$, is listed in table \ref{tab:normalization_density}. $\mu_0$ is the dielectric constant, and the central density $n_0$ is set to $0.76464 \cdot 10^{20} \, \mathrm{s}^{-3}$. Also, $\gamma$ is the specific heat ratio and equals $5/3$.

\par The resistivity is chosen to be the Spitzer resistivity with the dependency $\eta = \eta_0 \cdot (T/T_0)^{-3/2}$, where $\eta_0$ is calculated from the Spitzer formula \cite[7]{hoelzlJOREKNonlinearExtended2021} using the core temperature $T_0$ of the scenario. 
The same temperature dependency is applied also for the perpendicular viscosity for simplicity. 

\begin{table}
\centering
\caption{Normalization densities}
\begin{tabular}{l|l|l|l}
$A_\mathrm{eff}$ & 2.0 & 2.5 & 3.0 \\ \hline
$\rho_0$ & $2.5579 \cdot 10^{-7}$ & $3.1973 \cdot 10^{-7}$ & $3.8369 \cdot 10^{-7}$
\end{tabular}
\label{tab:normalization_density}
\end{table}

\par
\begin{figure}
  \center{\includegraphics{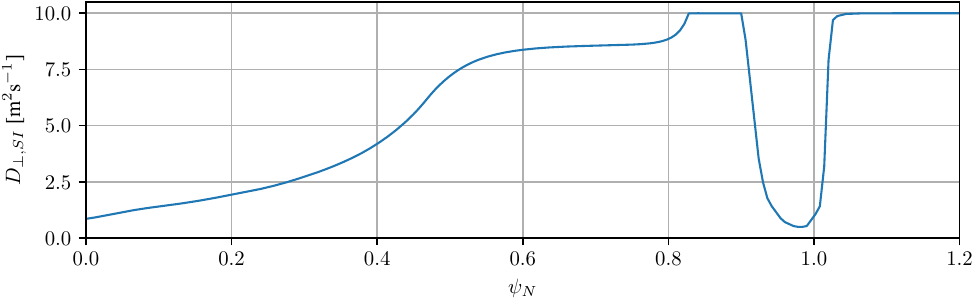}}
  \captionof{figure}{Perpendicular particle diffusivity input profile, which was matched by a previous work of A. Cathey \cite{catheyNonlinearExtendedMHD2020} to the experimental density and temperature profiles for D. Most important is the valley near the edge around $\psi_N = 1.0$, which is responsible for the build-up of the pedestal \cite{wagnerRegimeImprovedConfinement1982}.}
  \label{fig:D_perp}
\end{figure}

\begin{figure}
  \center{\includegraphics{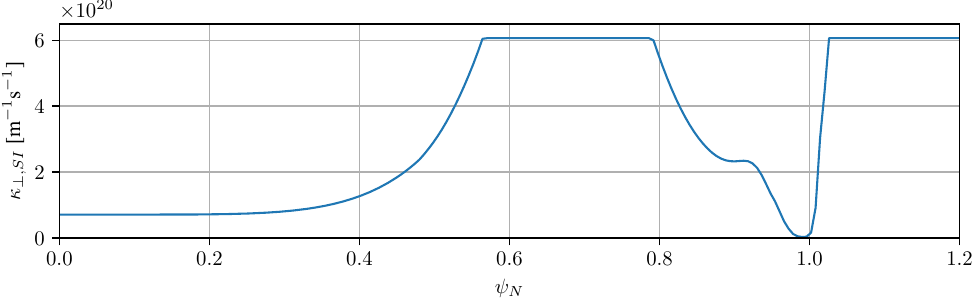}}
  \captionof{figure}{Perpendicular heat diffusivity input profile, matched to the experimental density and temperature profiles for D. Similarly as for the heat diffusivity, the valley near the edge around $\psi_N = 1.0$ represents the edge transport barrier.}
  \label{fig:kappa_perp}
\end{figure}

 Figures \ref{fig:D_perp} and \ref{fig:kappa_perp} visualize the diffusion profiles. Most important is the valley near the edge ($\psi_N \approx 1$), which represents the transport barrier that is responsible for the build-up of the pedestal \cite{wagnerRegimeImprovedConfinement1982}. The D simulations were set up in a previous work by A. Cathey \cite{catheyNonlinearExtendedMHD2020}, where the diffusivities, together with the heat source and particle source, were designed to evolve the density and temperature profiles towards the experimentally measured profiles in time-scales relevant for the typical inter-\acrshort{elm} evolution in \acrshort{aug}. The parallel heat diffusivity is adjusted in the simulation by the Spitzer-Härm formula $\kappa_{\parallel, \mathrm{actual}} = \kappa_{\parallel, 0} \cdot (T / T_0)^{5/2}$ from \cite{spitzerTransportPhenomena1953}. The diffusion profiles for D were kept for D-T and T, instead of applying the changes indicated by studies as discussed in section \ref{different_isotopes_fundamentals}. This is, because only effects coming from \acrshort{mhd} were meant to be included in the comparison.

\par
\begin{figure}
  \center{\includegraphics{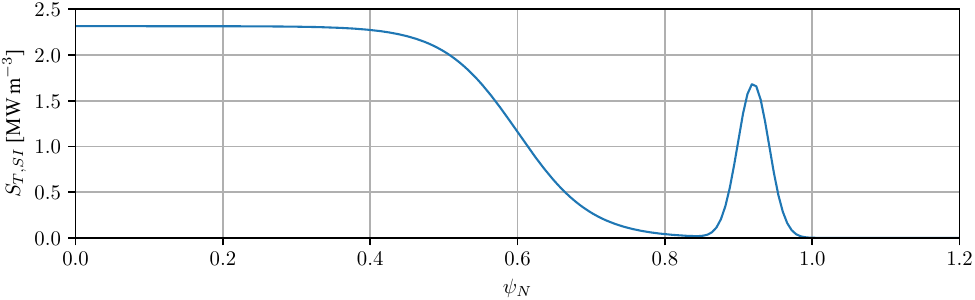}}
  \captionof{figure}{Heat source input profile, matched to the experimental density and temperature profiles for D. }
  \label{fig:heat_source}
\end{figure}

 Heating is modelled by the heat source profile $S_{E, SI}(\Psi_N)$. This is given as the sum of a tanh-shaped heat source, and a gaussian-shaped peak. The gaussian-shaped peak is positioned at $\psi_N = 0.92$ and has a width given by $\sigma = 0.03$, as visualized in figure \ref{fig:heat_source}.

\par
\begin{figure}
  \center{\includegraphics{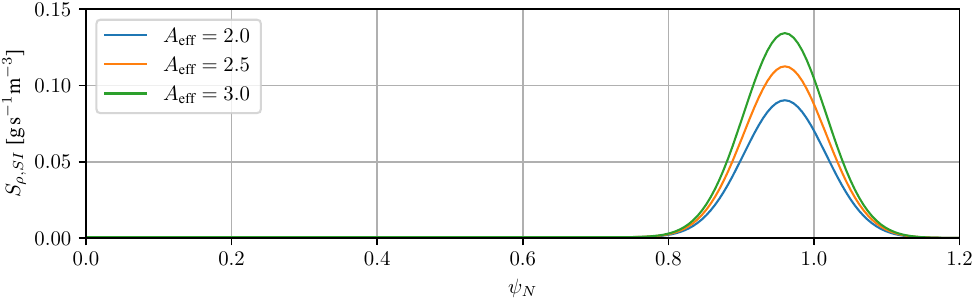}}
  \captionof{figure}{Particle source input profile, matched to the experimental density and temperature profiles for D. Here, the profiles for D-T ($A_\mathrm{eff} = 2.5$) and T ($A_\mathrm{eff} = 3.0$) were adjusted by a factor from equation \ref{eqn:particlesource_modification} to achieve the same particle density profiles for D-T and T as for D.}
  \label{fig:particle_source}
\end{figure}

 For the simulation, creation of new particles (e.g., by ionization) is modelled by defining a time-independent particle source. Its gaussian part has its center at $\psi_N = 0.96$ and a width given by $\sigma = 0.08$. It also has a tanh part, but this is very small compared to the gaussian part. This is visualized in figure \ref{fig:particle_source}. More explanation is given in the next subsection.

\subsection{Particle Source Modification}
\label{particle_source_modification}

\par If one would have used the same particle source profile for all normalized ion masses $A_\mathrm{eff}$, the resulting temperature and density would have been smaller, the higher the ion mass. This can be seen by rewriting equation (\ref{eqn:jorek_density}) in SI units:

\begin{equation}
\frac{\partial (m_{i, SI} n_{SI})}{\partial t_{SI}} = -\vec{\nabla} \cdot (m_{i, SI} n_{SI} \vec{V_{SI}}) + \vec{\nabla} \cdot (m_{i, SI} \underline{D_{SI}} \vec{\nabla} n_{SI}) + S_{\rho, SI}
\end{equation}

Therefore, to achieve the same particle density $n_{SI}$, $S_{\rho, SI}$ may not be constant but must be adjusted to the new isotope mass as follows:

\begin{equation}
\label{eqn:particlesource_modification}
S_{\rho, SI}' = S_{\rho, SI} \frac{A_\mathrm{eff}'}{A_\mathrm{eff}}
\end{equation}

\par The resulting profile for the three ion masses is shown in figure \ref{fig:particle_source}.

\subsection{Comparison between the Ion Mass Configurations}
\label{axisymmetric_comparison}

\begin{figure}
  \center{\includegraphics{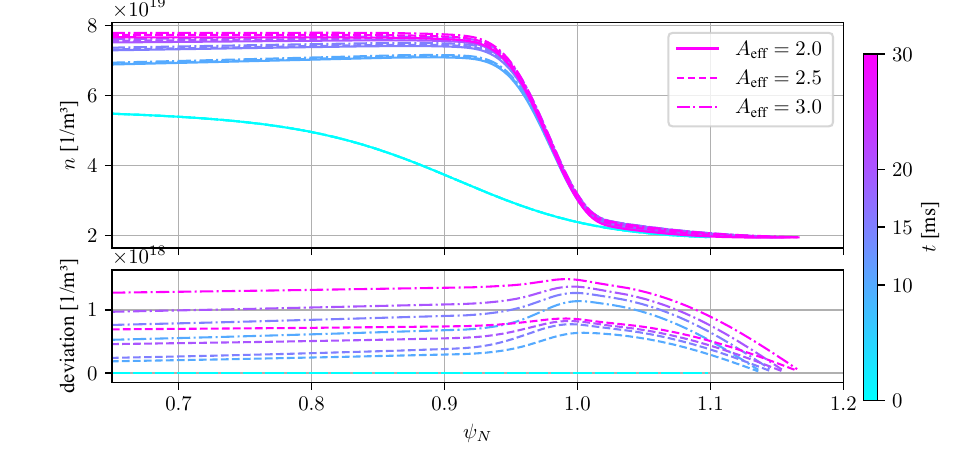}}
  \captionof{figure}{Particle density $n$ and its time evolution for the three normalized ion masses $A_\mathrm{eff}$, when using the adjusted particlesource as calculated from equation (\ref{eqn:particlesource_modification}). Shown profiles are for points lying on the midplane. The profile and the deviation between the ion masses is plotted for times 5, 10, 15, 20 and 30\,ms. The deviation is calculated by subtracting the density of $A_\mathrm{eff} = 2.0$ from the one of $A_\mathrm{eff} = 2.5$ and $A_\mathrm{eff} = 3.0$, respectively.}
  \label{fig:axisymmetric_density_psFromFormula}
\end{figure}

\par The simulation was first evolved axisymmetrically\footnote{One reason for why one cannot immediately start with non-axisymmetrical simulation, is that the velocity at open flux surfaces is set to ion sound speed. This is called Mach-1 Bohm boundary conditions, and requires to run first axisymmetrically for a short while to establish parallel flows in the \acrshort{sol} - see footnote 30 in \cite{hoelzlJOREKNonlinearExtended2021}.}, while comparing the profiles between the three ion mass configurations. The time evolution for the particle density is shown in figure \ref{fig:axisymmetric_density_psFromFormula}. For $t = 0$, the alignment is excellent, but for $t > 0$, small differences occur. This is also the case for the temperature profile, which is not shown here: The profiles from the initialization from the Grad-Shafranov balance equation match perfectly, whereas the differences are a result of the time evolution for $t > 0$.

\begin{figure}
  \center{\includegraphics{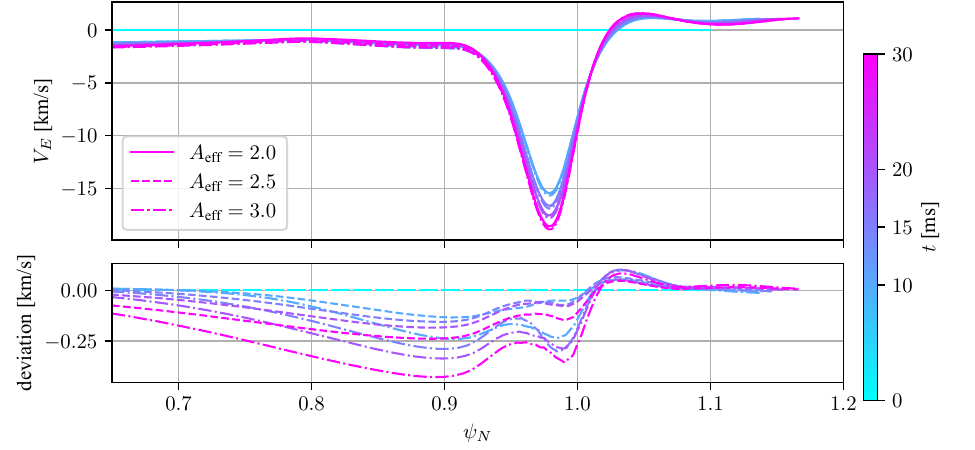}}
  \captionof{figure}{Time evolution of the ExB drift $\vec{v}_E$ in $\theta$-direction for the three normalized ion masses $A_\mathrm{eff}$, when using a slightly reduced particle source instead of the one given by equation \ref{eqn:particlesource_modification}. The profile and the deviation between the ion masses is plotted for times 5, 10, 15, 20 and 30\,ms. Shown profiles are for points lying on the outboard midplane. The deviation is calculated by subtracting the ExB drift of $A_\mathrm{eff} = 2.0$ from the one of $A_\mathrm{eff} = 2.5$ and $A_\mathrm{eff} = 3.0$, respectively.}
  \label{fig:ExB_velocity2}
\end{figure}

\par To reduce the deviations in the particle density between the ion masses, the particle sources were reduced by 0.3\% for $A_\mathrm{eff} = 2.5$ and 0.8\% for $A_\mathrm{eff} = 3.0$.
The deviations of other quantities due to this reduction changed only very slightly, or changed only in shape, but not in magnitude. 

\par The highest relative differences between the profiles for the different ion masses at the pedestal were recorded for the velocities. This is illustrated by the ExB-velocity, which is plotted in figure \ref{fig:ExB_velocity2}. 

\section{Single-n simulation: Unstable Modes}
\label{single_n_simulation}

\par In the simulations explained previously, the plasma was built up slowly over time, with the pressure gradient and current density increasing over time, thus acting destabilizing. Simultaneously, the ExB flows increased, providing stabilization through shearing \cite{pamelaInfluencePoloidalEquilibrium2010}. The interplay of these stabilizing and destabilizing terms leads to a non-trivial evolution of the growth rate spectrum over time. In \cite{catheyNonlinearMHDSimulations2021}, it had been shown that this interplay is also a key ingredient for understanding the explosive onset of the ELM crash.
Starting from multiple time points in this evolution of the $n=0$ equilibrium, JOREK was configured to only allow a selected toroidal harmonic $n$ additional to the axisymmetric evolution. This harmonic was initialized to noise level.
The unstable mode growth was then analyzed using the magnetic energies of the perturbations. The per-mode magnetic energy is calculated from the poloidal magnetic field corresponding to a given toroidal mode number $n$ as follows:

\begin{equation}
E_{\mathrm{mag}, n} = \int \frac{1}{2} \frac{\vec{B}_{\mathrm{pol}, n}^2}{\mu_0} \,dV
\end{equation}

\begin{figure}
  \center{\includegraphics{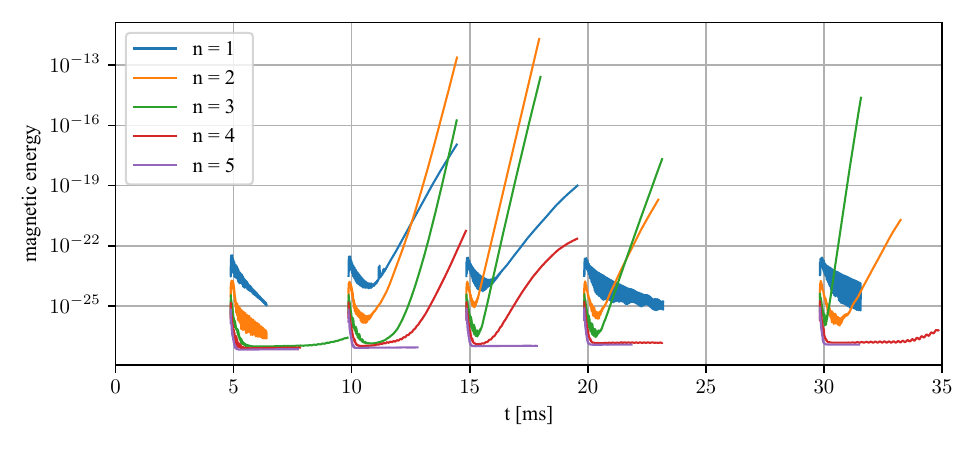}}
  \captionof{figure}{ Magnetic energies comparing the different unstable modes when starting to evolve a toroidal mode given by $n$ at a starting point of time around 5, 10, 15, 20 and 30\,ms. The modes with $n = 7$ and $n = 17$ were evolved too, but they show no instability, so that their energies behave as the one of $n = 5$. }
  \label{fig:magnetic_energies_singleN_m20}
\end{figure}

\par These simulations, which start off from the axisymmetric evolution to then continue with a single mode, were performed for the starting time points of 5, 10, 15, 20 and 30 ms. The magnetic energy's evolution is plotted in figure \ref{fig:magnetic_energies_singleN_m20}. Unstable modes can be recognized by their rising magnetic energy. It is important to note that, depending on $n$, the simulation around the noise level is not very precise due to numerical reasons, which is visible from the oscillative low-energy curves of $n=1$ during initialization and in case of stability. When the modes reach a certain energy threshold, the growth becomes more stable.

\par Modes $n = 5, 7$ and $17$ were found to be always stable. For $n = 1, 2, 3$ and $4$, stability was dependent on the time. The dominant poloidal components of these toroidal mode numbers had the poloidal mode number $m=7, 13, 19$ and $25$, respectively. The safety factor $q$ at the edge ($\psi_N \approx 0.97$) is 6.2. This does fit well to $q \approx \frac{m}{n}$ \cite[8]{cowleyLectureKink}: The expected $q$ would be 7 for $n = 1$, 6.5 for $n = 2$, 6.25 for $n=3$ and 6.33 for $n=4$.

\begin{figure}
  \center{\includegraphics[width=\textwidth]{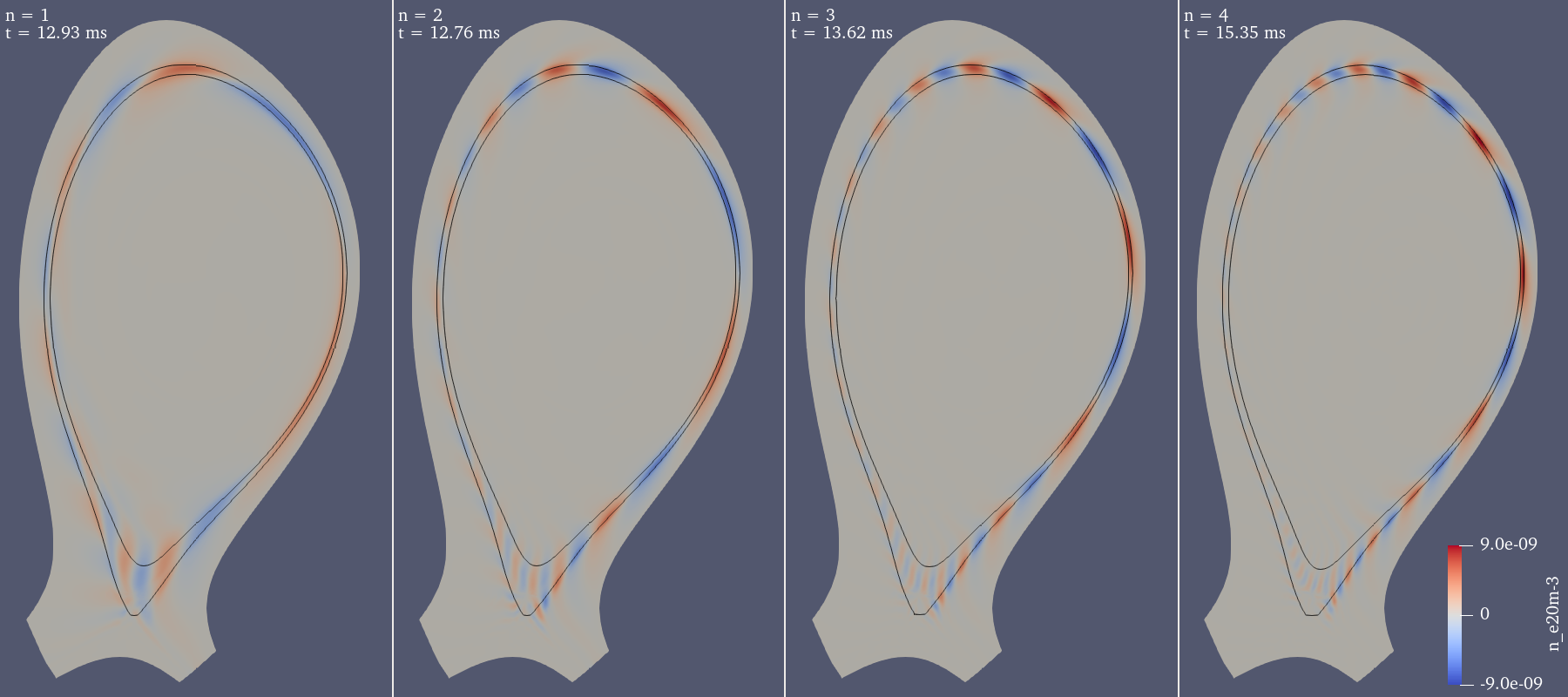}}
  \captionof{figure}{Comparison of the 4 modes found around 10 ms for $A_\mathrm{eff} = 2.0$ at same magnetic energy. The flux surfaces for $\psi_N = 0.96$ and $\psi_N = 1.0$ are indicated by a black line, each.}
  \label{fig:mode_comparison_rho_central_mass20}
\end{figure}

\par Figure \ref{fig:mode_comparison_rho_central_mass20} compares the structure of the four modes. All of them are modes located very close to the boundary of the confined plasma region, just inside the separatrix. For $n=1$, the particle density on the \acrshort{hfs} (left) has roughly the same strength as on the \acrshort{lfs}, indicating a peeling mode. The low poloidal mode number\footnote{The poloidal mode number can be determined by counting the maxima in $\theta$-direction, or by determining the highest-amplitude bin from a Fourier analysis at the edge.} $m$ is typical for peeling modes. Towards higher $n$, the \acrshort{hfs} becomes weaker, while the \acrshort{lfs} becomes stronger, indicating the increasing ballooning component of these modes. Over time, this structure does not change significantly.

\begin{figure}
  \center{\includegraphics{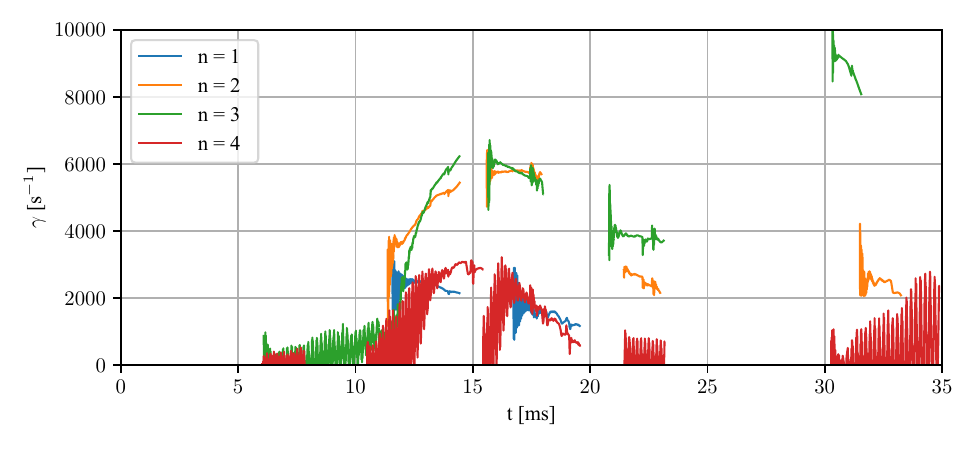}}
  \captionof{figure}{ Growth rates comparing the different unstable modes starting at $t = 5, 10, 15, 20$ and $30$\,ms, visualized by one line for each simulation. Modes with $n = 5$, $n = 7$ and $n = 17$ were also simulated, but were found to be linearly stable and are therefore not shown. }
  \label{fig:growth_rates_singleN_m20}
\end{figure}

\par The energy is proportional to the square of the perturbed magnetic field, and was therefore used as an indicator of the mode growth. The perturbations grow like $\exp{(\gamma t)}$, where $\gamma$ is the growth rate. By taking the derivative of the square root of the energy, the instantaneous growth rate was determined and plotted in figure \ref{fig:growth_rates_singleN_m20}. As especially the $n = 4$ mode grows quite slowly, it takes some time until it reaches a region of numerically stable growth rates. This becomes visible from the oscillative behavior of the red $n = 4$ curve in figure \ref{fig:growth_rates_singleN_m20}. In the next section, the initial growth rates will be determined using an exponential fit to get more precise results.

\begin{figure}
  \center{\includegraphics{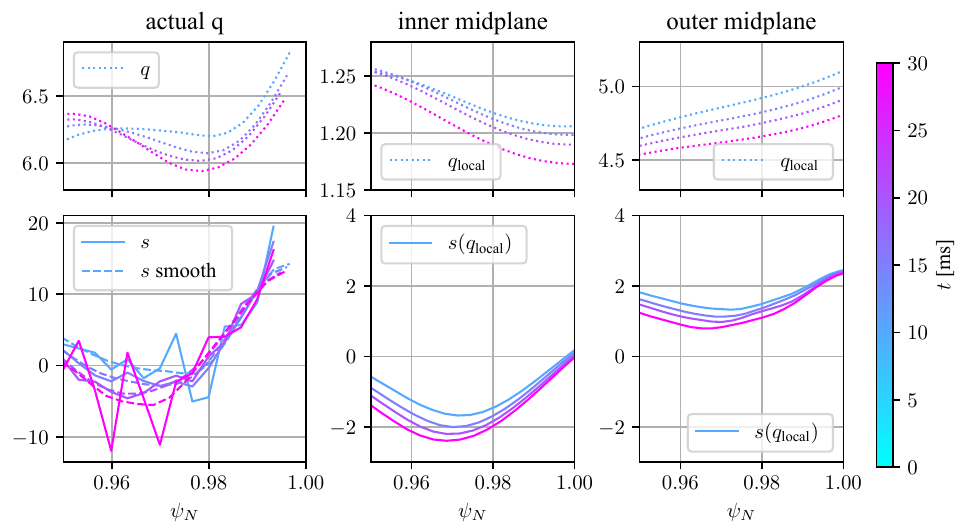}}
  \captionof{figure}{ Safety factor and magnetic shear at the edge for $t = 10, 15, 20$ and $30$\,ms. The actual safety factor $q$ (upper left diagram) is not very smooth due to numerics, causing numerical noise in the magnetic shear $s$ (lower left diagram). Smoothening $q$ yields also a smoother $s$ (dotted, lower left diagram). The center and right plots show the local $q$ (equation (\ref{eqn:safety_factor})) on the midplane, which is integrated poloidally over a flux surface to get the actual $q$. The local $q$ confirms the increase in negative shear over time indicated by the shear from the smoothened $q$. }
  \label{fig:magnetic_shear}
\end{figure}

\par Over time, the distance to the perfectly conducting wall increases slightly as indicated in figure \ref{fig:separatrix_evolution}, which should destabilize ballooning modes \cite[302]{freidbergPlasmaPhysicsFusion2007}. Also, the pressure gradient increases, which should have a destabilizing influence on ballooning modes as well. While this does at least fit roughly to the behavior of the $n=3$ mode, it cannot explain the reduced growth rate of all modes from 20 to 25 ms. This could be explained by taking into account the magnetic shear $s = r/q \, \mathrm{d} q / \mathrm{d} r$ with the safety factor $q$ (see section \ref{diamagnetic_current}). Both high positive magnetic shear (see \cite{zohmMagnetohydrodynamicStabilityTokamaks} chapter 5), as well as high negative magnetic shear (derived in \cite{antonsenPhysicalMechanismEnhanced1996}) stabilizes ballooning modes, because this makes it more difficult for the mode to expand radially. The output safety factor of JOREK is quite noisy due to numerical issues. By smoothening\footnote{using the function \texttt{scipy.interpolate.make\_smoothing\_spline}} $q$, one can achieve also a smoother $s$, shown dotted in the left diagram of figure \ref{fig:magnetic_shear}. The major change over time is that due to the increasing bootstrap current, the shear becomes more negative in the region of $\psi_N = 0.96..0.98$, which acts stabilizing on the mode.

\par It is far easier to calculate $q_\mathrm{local} = r B_\phi / (R B_\theta)$ at a specific position, whose average over $\theta$ along a flux surface represents the actual $q$ as shown in equation \ref{eqn:safety_factor}. The local $q$ was evaluated along the inner and outer midplane, resulting in a smaller local $q$ than the actual $q$, as shown in figure \ref{fig:magnetic_shear}. This is because of smaller $r$ at the midplane, compared to the upper and lower end of the plasma when regarding positions along the same fluxsurface. The shear $s$ calculated from the local $q$ behaves similarly as the one corresponding to the total $q$, thus confirming that smoothening $q$ did not lead to wrong conclusions regarding $s$. 

\par Another stabilizing factor are sheared flows of the plasma due to the ExB drift \cite{pamelaInfluencePoloidalEquilibrium2010} (plotted in figure \ref{fig:ExB_velocity2}), as well as the ion diamagnetic drift \cite[7]{snyderEdgeLocalizedModes2002}. Both increase over time due to the increasing pressure gradient. It is conceivable that the interplay of stabilization due to the VxE drift, the diamagnetic velocity, and negative magnetic shear with the destabilization due to increasing pressure gradient and the increased distance to the wall can lead to stabilization as well as destabilization (dependent on which effect dominates). This could explain the reduction and increase of the growth rate of the modes with a stronger ballooning component, which are those with $n = 2, 3$ and $4$, over time.

\par Peeling modes like $n = 1$ are destabilized by the edge current density gradient $\nabla j_\phi$, and stabilized by the edge pressure gradient \cite{connorMagnetohydrodynamicStabilityTokamak1998}. Here, both are rising, with the growth being very similar, because the bootstrap current links both of them. An additional stabilizing influence comes from the shear of the ExB flow. Again, the complex interplay of these influences is expected to the reason for why the $n=1$ mode becomes stable.

\subsection{Single-n Growth Rate Comparison Between Ion Masses}
\label{single_n_growth_rate_comparison}

\par
\begin{figure}
  \center{\includegraphics{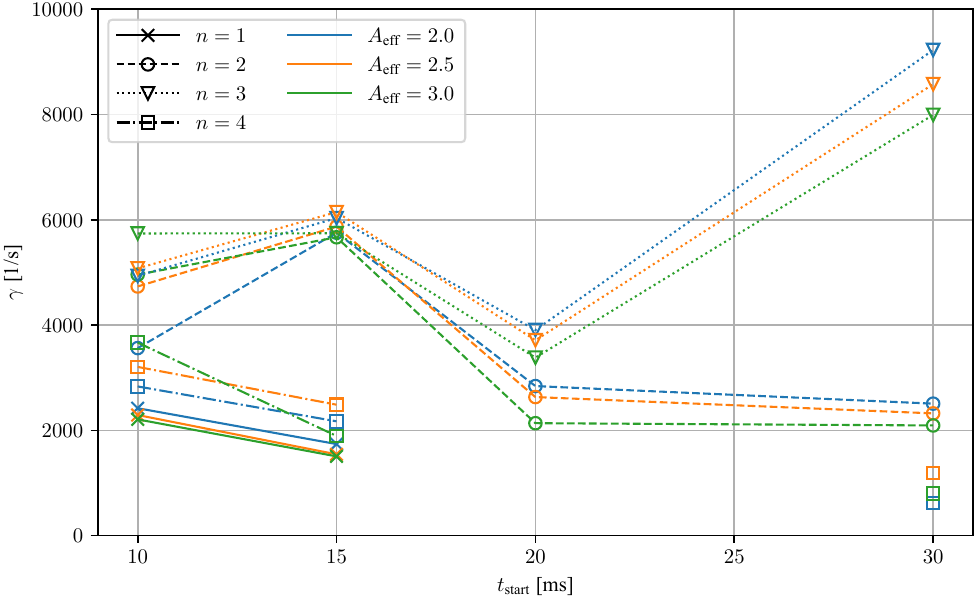}}
  \captionof{figure}{ Growth rates for simulations with one toroidal mode number $n$, additionally to the axisymmetric mode $n = 0$, for the three simulated ion masses $A_\mathrm{eff} = 2.0$, $2.5$ and $3.0$. Between the different start times $t_\mathrm{start}$, the pedestals evolved, which caused changes in the growth rates.}
  \label{fig:fitted_growth_rates_vs_t}
\end{figure}

 The simulations run in the previous section for $A_\mathrm{eff} = 2.0$ were now also run for the other two ion mass configurations. Figures \ref{fig:fitted_growth_rates_vs_t} compares the growth rates between the ion masses. They were determined using an exponential fit of the magnetic energy, taking only into account the data near the starting time point of the perturbation (ignoring further changes in the growth rate incl. non-linear behavior)\footnote{A simple graphical user interface was created, that allows to select the start and end point of the fit for each ion mass and starting time point, so that the period of the initial growth can be specified by the user. It then calculates the fit using \texttt{scipy.stats.linregress} applied on the logarithm of the magnetic energy, and finally displays the fitted curve in the magnetic energy plots.}. The growth rates determined in this way are also printed in tables \ref{tab:growth_rates_linear_10ms} to \ref{tab:growth_rates_linear_30ms}. For 5\,ms, all modes turned out to be stable, therefore, this time was not included in figure \ref{fig:fitted_growth_rates_vs_t} and in tables.

\begin{table}
\centering
\caption{Growth rates for single-n simulations, start at 10\,ms}
\begin{tabular}{l|l|l|l|l|l|l|l|l|l|l}
 & \multicolumn{2}{c|}{$n = 1$} & \multicolumn{2}{c|}{$n = 2$} & \multicolumn{2}{c|}{$n = 3$} & \multicolumn{2}{c|}{$n = 4$} & \multicolumn{2}{c}{$n = 5, 7, 17$} \\ \cline{2-11}
$A_\mathrm{eff}$ & $\gamma$ [1/s] & $m$ & $\gamma$ [1/s] & $m$ & $\gamma$ [1/s] & $m$ & $\gamma$ [1/s] & $m$ & $\gamma$ [1/s] & $m$ \\ \hline
\hline
2.0 & 2420 & 7 & 3560 & 13 & 4940 & 19 & 2830 & 25 & - & - \\ \hline
2.5 & 2290 & 7 & 4730 & 13 & 5080 & 19 & 3210 & 25 & - & - \\ \hline
3.0 & 2210 & 7 & 4960 & 13 & 5740 & 19 & 3670 & 25 & - & -
\end{tabular}
\label{tab:growth_rates_linear_10ms}
\end{table}

\begin{table}
\centering
\caption{Growth rates for single-n simulations, start at 15\,ms}
\begin{tabular}{l|l|l|l|l|l|l|l|l|l|l}
 & \multicolumn{2}{c|}{$n = 1$} & \multicolumn{2}{c|}{$n = 2$} & \multicolumn{2}{c|}{$n = 3$} & \multicolumn{2}{c|}{$n = 4$} & \multicolumn{2}{c}{$n = 5, 7, 17$} \\ \cline{2-11}
$A_\mathrm{eff}$ & $\gamma$ [1/s] & $m$ & $\gamma$ [1/s] & $m$ & $\gamma$ [1/s] & $m$ & $\gamma$ [1/s] & $m$ & $\gamma$ [1/s] & $m$ \\ \hline
\hline
2.0 & 1740 & 7 & 5760 & 13 & 6030 & 19 & 2170 & 25 & - & - \\ \hline
2.5 & 1540 & 7 & 5870 & 13 & 6150 & 19 & 2490 & 25 & - & - \\ \hline
3.0 & 1510 & 7 & 5670 & 13 & 5750 & 19 & 1890 & 25 & - & -
\end{tabular}
\label{tab:growth_rates_linear_15ms}
\end{table}

\begin{table}
\centering
\caption{Growth rates for single-n simulations, start at 20\,ms}
\begin{tabular}{l|l|l|l|l|l|l|l|l|l|l}
 & \multicolumn{2}{c|}{$n = 1$} & \multicolumn{2}{c|}{$n = 2$} & \multicolumn{2}{c|}{$n = 3$} & \multicolumn{2}{c|}{$n = 4$} & \multicolumn{2}{c}{$n = 5, 7, 17$} \\ \cline{2-11}
$A_\mathrm{eff}$ & $\gamma$ [1/s] & $m$ & $\gamma$ [1/s] & $m$ & $\gamma$ [1/s] & $m$ & $\gamma$ [1/s] & $m$ & $\gamma$ [1/s] & $m$ \\ \hline
\hline
2.0 & - & - & 2840 & 13 & 3910 & 19 & - & - & - & - \\ \hline
2.5 & - & - & 2630 & 13 & 3710 & 19 & - & - & - & - \\ \hline
3.0 & - & - & 2140 & 13 & 3380 & 19 & - & - & - & - 
\end{tabular}
\label{tab:growth_rates_linear_20ms}
\end{table}

\begin{table}
\centering
\caption{Growth rates for single-n simulations, start at 30\,ms}
\begin{tabular}{l|l|l|l|l|l|l|l|l|l|l}
 & \multicolumn{2}{c|}{$n = 1$} & \multicolumn{2}{c|}{$n = 2$} & \multicolumn{2}{c|}{$n = 3$} & \multicolumn{2}{c|}{$n = 4$} & \multicolumn{2}{c}{$n = 5, 7, 17$} \\ \cline{2-11}
$A_\mathrm{eff}$ & $\gamma$ [1/s] & $m$ & $\gamma$ [1/s] & $m$ & $\gamma$ [1/s] & $m$ & $\gamma$ [1/s] & $m$ & $\gamma$ [1/s] & $m$ \\ \hline
\hline
2.0 & - & - & 2510 & 13 & 9230 & 19 & 620 & 25 & - & - \\ \hline
2.5 & - & - & 2320 & 13 & 8580 & 19 & 1190 & 25 & - & - \\ \hline
3.0 & - & - & 2090 & 13 & 7990 & 19 & 800 & 25 & - & -
\end{tabular}
\label{tab:growth_rates_linear_30ms}
\end{table}

\par The poloidal mode number did not differ across ion masses. The relative difference of the growth rates between the configurations is quite small, contradicting the stabilization related to the ion diamagnetic frequency discussed in chapter \ref{different_isotopes_fundamentals} from \cite[13]{horvathIsotopeDependenceType2021}. It is likely that this is, because there, the ion diamagnetic velocity was not explicitly included in the simulation, but only added retrospectively when assessing the growth rates. Many more simplifications were applied in the referred model \cite{huijsmansDiamagneticStabilization2001}, such as ignoring the magnetic curvature. The model used in this thesis does not employ most of these simplifications, and incorporates the ion diamagnetic velocity directly in the simulation. This renders a direct comparison invalid.

\par It is visible that for $n > 1$, at 10\,ms, the growth rates are highest for $A_\mathrm{eff} = 2.0$, while for 15\,ms, they are highest for $A_\mathrm{eff} = 2.5$, and for 20\,ms, they are highest for $A_\mathrm{eff} = 3.0$. In contrast to that, the $n = 1$ mode grows always fastest for $A_\mathrm{eff} = 2.0$, if present at all. The difference in the deviation of the $n=1$ mode compared to the $n = 2, 3, 4$ modes might be related to the fact the former is mostly peeling-like, while the latter have a stronger ballooning component.

\subsection{Momentum Equation}
\label{momentum_equation_explanation}

\par To gain insight into the differences of the axisymmetric quantities shown in section \ref{axisymmetric_comparison}, as well as into the differences in the growth rates shown in section \ref{single_n_growth_rate_comparison}, the momentum equation was considered. The reason is that (apart from the small modification of the particle source) the momentum equation is the only \acrshort{mhd} equation which is modified by more than a simple factor. In the following, the equation specific to the perpendicular part $\vec{v}_E$ is given, as implemented in the JOREK code (with equation \ref{eqn:particlesource_modification} plugged in)\footnote{\cite{hoelzlJOREKNonlinearExtended2021} also includes a term $(\nabla \cdot (D_{SI} \nabla n_{SI})) \vec{v}_{E, SI}$, but this is not implemented in the specific model used in this thesis.}:

\begin{multline}
\rho_{SI} \frac{\mathrm{d} \vec{v}_{E, SI}}{\mathrm{d} t_{SI}} = - \rho_{SI} \vec{v}_{i, SI} \cdot \nabla \vec{v}_{E, SI} - \vec{\nabla}_\perp p_{SI} + \vec{J}_{SI} \times \vec{B}_{SI} + \mu_{SI} \nabla^2 (\vec{v}_{E, SI} + \vec{v}_{i, SI})
\\ + \mu_{hyp, SI} \nabla^4 (\vec{v}_{E, SI} + \vec{v}_{i, SI}) - \frac{A_\mathrm{eff}}{2} S_{\rho, SI} \vec{v}_{E, SI}
\end{multline} 

\par The ion diamagnetic velocity is $\vec{v}_{i, SI} = \frac{1}{e \, n_{SI}} \frac{\vec{B}_{\phi, SI} \times \vec{p}_{SI}}{B_{\phi, SI}^2} \frac{1}{2}$, using the definition of $\delta^\star$ as given in section \ref{approach_jorek} and using $\vec{B}_\phi = B_0 \frac{R_0}{R} \vec{e}_\phi$, which is constant in time. The tensor product and dyad definition can be used to rewrite $\vec{u} \cdot \nabla \vec{v} = \Sigma^3_{i = 1} \Sigma^3_{j = 1} \vec{e}_i u_j \frac{\partial v_i}{\partial x_j}$ (see chapter 2 of \cite{schnackLecturesMHD2009}). Here, $u_i$ is the $i$\textsuperscript{th} component of the vector $\vec{u}$. The total time derivate of the ExB velocity separates to ${\frac{\mathrm{d}\vec{v}_{E, SI}}{\mathrm{d} t} = \frac{\partial \vec{v}_{E, SI}}{\partial t} + \vec{v}_{E, SI} \cdot \nabla \vec{v}_{E, SI}}$.

\par Some parts of the equation scale with the mass density, and therefore also with the ion mass, while others stay constant. The effect of this was investigated by an artificial test, for which a factor $f = 2 / A_\mathrm{eff}$ was included in all density-related terms as follows:

\begin{multline}
\label{eqn:modified_momentum_equation}
f \rho_{SI} \frac{\mathrm{d} \vec{v}_{E, SI}}{\mathrm{d} t_{SI}} = - f \rho_{SI} \vec{v}_{i, SI} \cdot \nabla \vec{v}_{E, SI} - \vec{\nabla}_\perp p_{SI} + \vec{J}_{SI} \times \vec{B}_{SI} + \mu_{SI} \nabla^2 (\vec{v}_{E, SI} + \vec{v}_{i, SI})
\\ + \mu_{hyp, SI} \nabla^4 (\vec{v}_{E, SI} + \vec{v}_{i, SI}) - f \frac{A_\mathrm{eff}}{2} S_{\rho, SI} \vec{v}_{E, SI}
\end{multline} 

\par This will make the equation be the same (in SI units), independently of the ion mass. For the parallel velocity equation, this was done similarly. Of course, the resulting modified momentum equation is no longer correct with respect to the physical reality.

\begin{figure}
  \center{\includegraphics{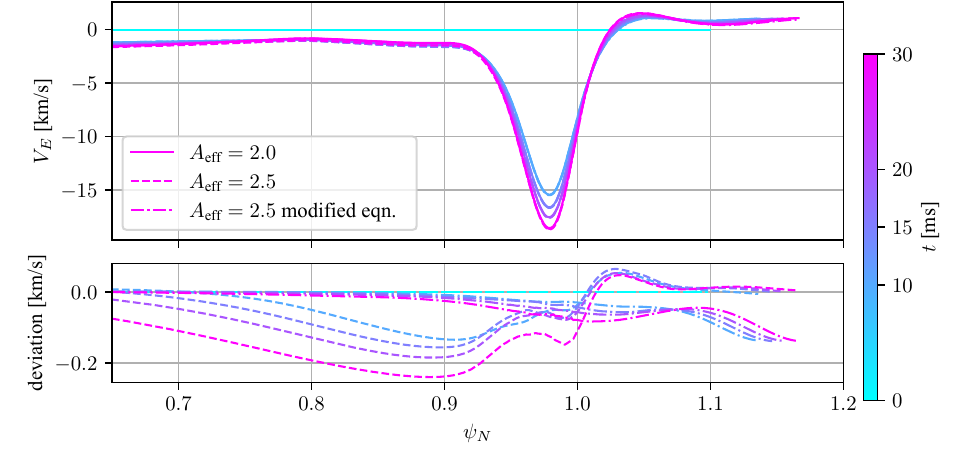}}
  \captionof{figure}{ExB drift $\vec{v}_E$ (perpendicular component parallel to the flux surface) and its time evolution, comparing the unmodified simulation with the simulation that removes the ion mass dependence from the momentum equation as shown in equation \ref{eqn:modified_momentum_equation}. }
  \label{fig:ExB_velocity_rhoCorrV4}
\end{figure}

\par When using the modified momentum equation, the overall profiles align better. Figure \ref{fig:ExB_velocity_rhoCorrV4} shows this for the perpendicular ExB velocity. As an exception to the overall improvement in similarity, in the \acrshort{sol}, the velocity difference to the $A_\mathrm{eff}=2.0$ case increases slightly due to the modification, but this should have only a negligible impact.

\par It is important to note that now, the exact particle source as demanded by formula (\ref{eqn:particlesource_modification}) is used. The slight adjustment of this value, which was used before to get more similar densities, is no longer necessary. It was verified that this slight adjustment has no or almost no influence on the growth rates, compared to the changes caused by the modification of the momentum equation.

\begin{figure}
  \center{\includegraphics{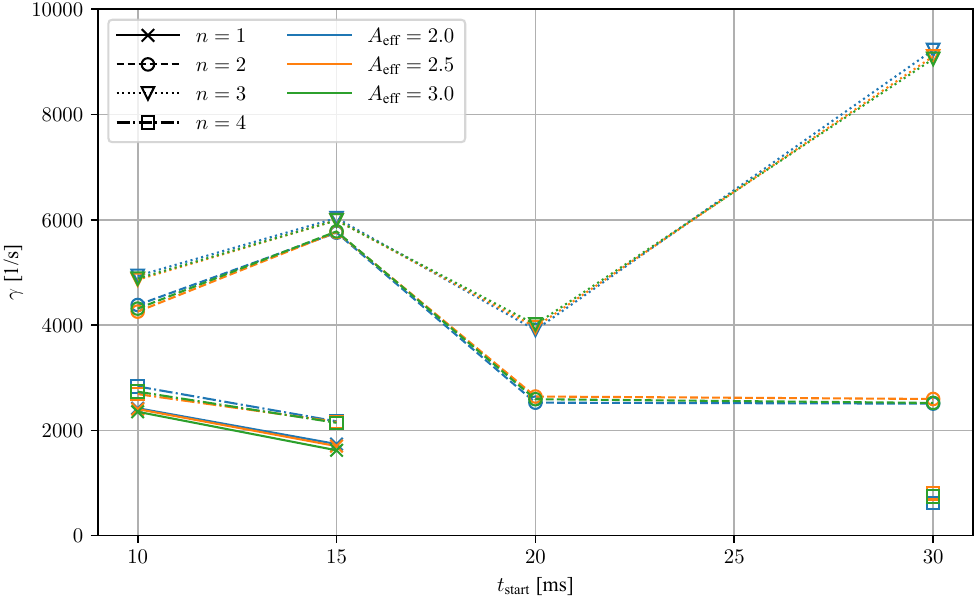}}
  \captionof{figure}{ Growth rates for simulations with one toroidal mode number $n$, additionally to the axisymmetric mode $n = 0$, for the three simulated ion masses $A_\mathrm{eff} = 2.0$, $2.5$ and $3.0$. The modified momentum equation (\ref{eqn:modified_momentum_equation}) was used, in which the terms proportional to the ion mass were scaled down so that they become constant with respect to the ion mass. This modification is not physically correct, but it shows that the proportional terms of the (physically correct) momentum equation used previously are the major reason for the changes in growth rates seen before.} 
  \label{fig:fitted_growth_rates_vs_t_rhoCorrV4}
\end{figure}

\par The resulting growth rates are compared in figure \ref{fig:fitted_growth_rates_vs_t_rhoCorrV4}. For all unstable modes, the alignment improves a lot (compare for figure \ref{fig:fitted_growth_rates_vs_t}), confirming that the combination of terms proportional to $A_\mathrm{eff}$ with terms constant with respect to $A_\mathrm{eff}$ in the momentum equation is indeed the source of the deviations between ion masses. Only a very small difference remains.

\begin{sloppypar} Further understanding can be gained by splitting the profiles into an e\-qui\-li\-bri\-um part (with ${\mathrm{d}\vec{v}_{E, SI} / \mathrm{d} t_{SI} = 0}$), which is denoted with subscript 0, and a small perturbed part $\propto \xi$, which is denoted with subscript 1. By dropping all parts which are $\mathcal{O}(\xi^2)$, one gets an equation for the equilibrium (SI subscript dropped for readability)
\end{sloppypar}

\begin{multline}
\label{eqn:momentum_equation_si_equilibrium}
0 = - \rho_0 \vec{v}_{i, 0} \cdot \nabla \vec{v}_{E, 0} - \vec{\nabla}_\perp p_0 + \vec{J}_0 \times \vec{B}_0 + \mu_0 \nabla^2 (\vec{v}_{E, 0} + \vec{v}_{i, 0})
\\ + \mu_{hyp, 0} \nabla^4 (\vec{v}_{E, 0} + \vec{v}_{i, 0}) - \frac{A_\mathrm{eff}}{2} S_{\rho, 0} \vec{v}_{E, 0}
\mathrm{,}
\end{multline} 

and one for the perturbation

\begin{multline}
\label{eqn:momentum_equation_si_perturbed}
\rho_0 \frac{\mathrm{d} \vec{v}_{E, 1}}{\mathrm{d} t} = - \rho_0 \vec{v}_{i, 0} \cdot \nabla \vec{v}_{E, 1} - \rho_0 \vec{v}_{i, 1} \cdot \nabla \vec{v}_{E, 0} - \rho_1 \vec{v}_{i, 0} \cdot \nabla \vec{v}_{E, 0} - \vec{\nabla}_\perp p_1 + \vec{J}_0 \times \vec{B}_1 + \vec{J}_1 \times \vec{B}_0 + \mu \nabla^2 (\vec{v}_{E, 1} + \vec{v}_{i, 1})
\\ + \mu_{hyp} \nabla^4 (\vec{v}_{E, 1} + \vec{v}_{i, 1}) - \frac{A_\mathrm{eff}}{2} S_\rho \vec{v}_{E, 1}
\end{multline} 

with $\vec{v}_{i, 1} = \frac{1}{e \, n_0 \, B^2_\phi} \left( \vec{B}_\phi \times \vec{p}_1 - \frac{n_1}{n_0} \vec{B}_\phi \times \vec{p}_0 \right) \frac{1}{2}$.

\par In JOREK, one can employ the $n = 0$ profiles of one configuration in another configuration, which was used here to apply the modification factor $f$ only on the perturbed equation (\ref{eqn:momentum_equation_si_perturbed}), or only on the $n=0$ equilibrium defined by equation (\ref{eqn:momentum_equation_si_equilibrium}). The resulting growth rates are shown in tables \ref{tab:growth_rates_n0const_n1} and \ref{tab:growth_rates_n0const_n3} (note that the growth rates are constant in time, because the equilibrium is kept constant, and because the perturbations are small compared to the equilibrium). For $n = 1$, changing from the unmodified to the modified equilibrium increases the growth rate by around 50\,1/s, whereas modifying the perturbation equation increases the growth rate by around 110\,1/s. For $n = 3$, the equilibrium change contributes around 440\,1/s, whereas the perturbation equation change contributes around 880\,1/s. Although the equilibrium changes in most profiles are very small (e.g., only a slight reduction of the maximum pressure gradient by 0.4\%), the variations in ExB velocity and ion diamagnetic drift are probably the reason for why modifying the equilibrium equation alone already affects the growth rates. But the major cause for the differences between ion masses is that the perturbations' evolution is affected through the momentum equation, as shown by the higher growth rates differences.

\begin{table}
\centering
\caption{Linear growth rates (in $1/\mathrm{s}$) dependence the momentum equations' modification on the axisymmetric $n = 0$ mode and on the perturbed mode $n = 1$, for $t = 12$\,ms, $A_\mathrm{eff} = 2.5$, compared to a growth rate of 2410 1/s for $A_\mathrm{eff} = 2.0$}
\begin{tabular}{l||r|r}
& $n = 1$ original & $n=1$ with modification \\  \hline \hline
$n = 0$ original & 2250 & 2360 \\ \hline
$n = 0$ with modification & 2300 & 2420
\end{tabular}
\label{tab:growth_rates_n0const_n1}
\end{table}

\begin{table}
\centering
\caption{Linear growth rates (in $1/\mathrm{s}$) dependence the momentum equations' modification on the axisymmetric $n = 0$ mode and on the perturbed mode $n = 3$, for $t = 12$\,ms, $A_\mathrm{eff} = 2.5$, compared to a growth rate of 2110\,1/s for $A_\mathrm{eff} = 2.0$}
\begin{tabular}{l||r|r}
& $n = 3$ original & $n=3$ with modification \\  \hline \hline
$n = 0$ original & 3480 & 2610 \\ \hline
$n = 0$ with modification & 3050 & 2160
\end{tabular}
\label{tab:growth_rates_n0const_n3}
\end{table}

\par The main point of this section is that the difference in growth rates between the simulations can mostly be explained by the $\rho$-dependence (and therefore $A_\mathrm{eff}$-dependence) of some terms in the momentum equation. When employing the modified (non-physical) momentum equation, all equations of the resulting model are the same in SI units (or only differ by a common factor of the right and left hand side). This means, that the remaining (slight) differences between the ion mass configurations cannot come from \acrshort{mhd}.
One source might be, that the boundary conditions were not considered, and might cause an ion mass dependence.

\section{Multi-n Simulation}
\label{multi_n_simulation}

\par
\begin{figure}
  \center{\includegraphics{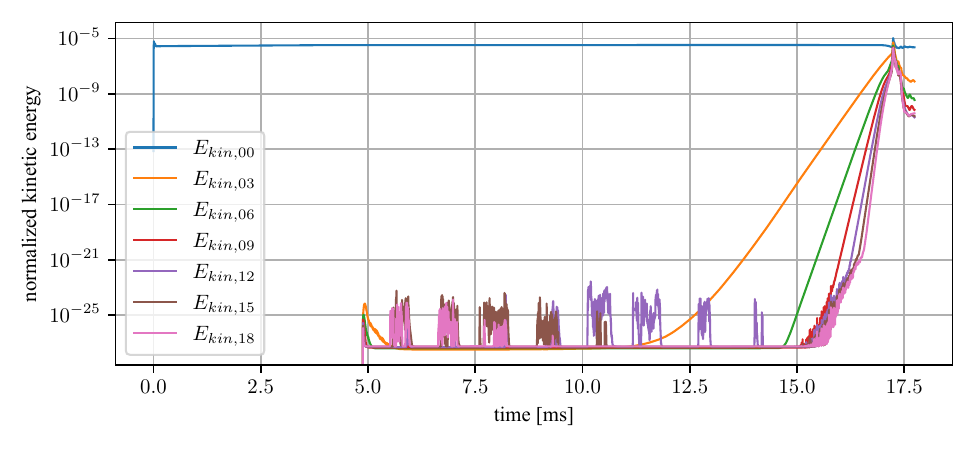}}
  \captionof{figure}{ Kinetic energy evolution for normalized ion mass of 2.0 that was run with toroidal modes $n = 0, 3, 6, 9, 12, 15$ and $18$, preceded by an axisymmetric simulation until 5\,ms. Around $t = 17.2$\,ms, the peak of an \acrshort{elm} crash is seen, during which 7.2\% of the thermal energy and 3.0\% of the particles are lost. The smaller bumps in modes 12, 15 and 18 before 14\,ms are numerical noise.
  }
  \label{fig:kinetic_energies_m20}
\end{figure}

 As a last step, a simulation with multiple toroidal modes was performed: The toroidal modes 3, 6, 9, 12, 15 and 18 were incorporated, together with the axisymmetric mode. This allows coupling between modes, thus resulting in \acrshort{elm} simulations. Figure \ref{fig:kinetic_energies_m20} plots the kinetic energies of the modes. First of all, it must be noted that in modes 12, 15 and 18, smaller bumps occur at the beginning up to a normalized kinetic energy of around $10^{-23}$ which are numerical noise.

\par
\begin{figure}
  \center{\includegraphics{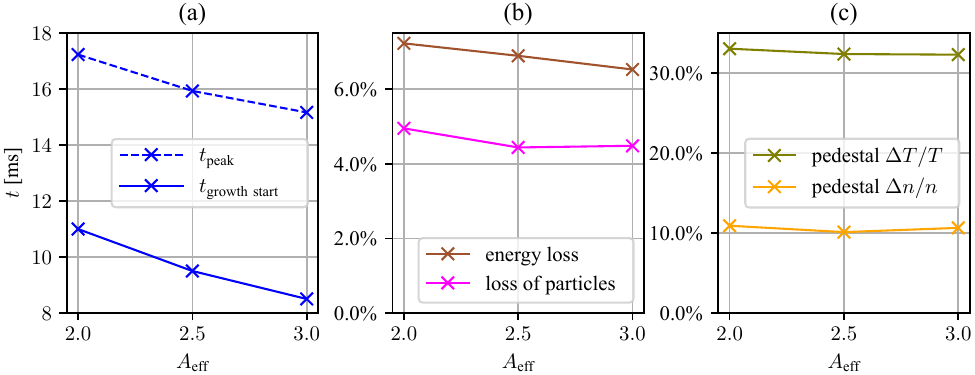}}
  \captionof{figure}{ Isotope comparison of the multi-n (\acrshort{elm}) simulation with toroidal modes $n = 0, 3, 6, 9, 12, 15$ and $18$. Plot (a) shows the shift of the crash towards earlier times for higher ion mass, which occurs because the $n = 3$ mode, which triggers the \acrshort{elm}, becomes unstable earlier for higher ion masses. $t_\mathrm{peak}$ was determined from the maximum of the kinetic energy of the non-axisymmetric modes, and $t_\mathrm{end}$ is the point of time when the thermal energy starts growing again, indicating the end of the ELM crash. Plot (b) shows the relative loss of the total energy and total particle content. Plot (c) was derived by extracting the pedestal height at $\psi_N = 0.9$.}
  \label{fig:elm_parameters}
\end{figure}

 After a period of linear growth in the $n = 3$ mode, the mode with $n = 6$ is non-linearly driven by the $n=3$ mode, which can be considered as a special case of three-wave interactions \cite{krebsNonlinearExcitationLown2013}, as it was found already for the D simulation in \cite[4]{catheyNonlinearExtendedMHD2020}: In general three-wave interaction, two modes with toroidal mode number $n_1$ and and $n_2$ can drive a third mode with $n_\mathrm{driven} = n_1 \pm n_2$, resulting in a growth rate of the driven mode $\gamma_\mathrm{driven} = \gamma_1 + \gamma_2$. Here, $n_1 = n_2 = 3$, so that the growth rate of the $n = 6$ mode would be twice the one of the $n = 3$ mode. This fits to the simulation, whose growth rates are $\gamma_3 \approx 5100$\,1/s and $\gamma_6 \approx 10100$\,1/s. Towards the maximum, more and more modes are driven until at the maximum, all toroidal modes are excited. Then, the simulation's thermal energy and particle content drops, indicating the release of heat and particles as is the case during an \acrshort{elm} crash. The losses are listed in table \ref{tab:multi_n_quantities} and plotted in figure \ref{fig:elm_parameters}.

\begin{table}
\centering
\caption{ELM parameters of the multi-n simulation with toroidal modes $n = 0, 3, 6, 9, 12, 15$ and $18$, compared between the ion masses $A_\mathrm{eff}$, as visualized in figure \ref{fig:elm_parameters}.}
\begin{tabular}{r|r|r|r|r|r|r|r}
normalized & growth & maximum's & $\Delta t$ & relative & relative & pedestal & pedestal \\
ion mass $A_\mathrm{eff}$ & start [ms] & time [ms] & [ms] & energy loss & particle loss & $\Delta T / T$ & $\Delta n / n$ \\ \hline \hline
2.0 & 11.0 & 17.2 & 6.2 & 7.2\% & 5.0\% & 33.1\% & 10.9\% \\ \hline
2.5 & 9.5 & 15.9 & 6.4 & 6.9\% & 4.4\% & 32.4\% & 10.1\% \\ \hline
3.0 & 8.5 & 15.2 & 6.7 & 6.5\% & 4.5\% & 32.3\% & 10.7\%
\end{tabular}
\label{tab:multi_n_quantities}
\end{table}

\begin{figure}
  \center{\includegraphics{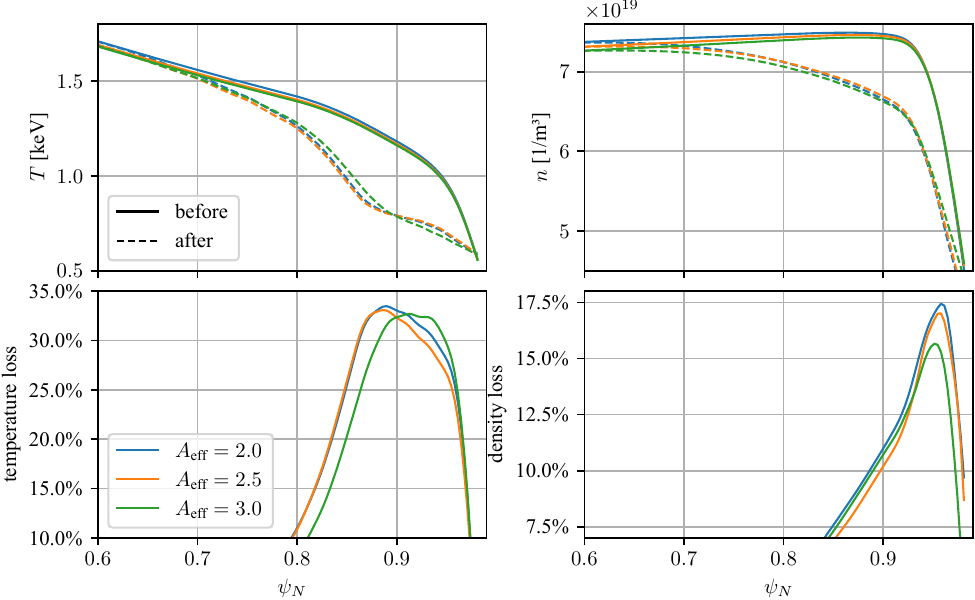}}
  \captionof{figure}{ Temperature (upper left) and density (upper right) profiles before and after the ELM crash, compared between the ion masses. These profiles are averages along a flux surface. The profiles' relative reductions (lower plots) are shown as well, which where calculated with ${(T_\mathrm{before} - T_\mathrm{after}) / T_\mathrm{before}}$ and ${(n_\mathrm{before} - n_\mathrm{after}) / n_\mathrm{before}}$ for the temperature and density, respectively. }
  \label{fig:elm_profiles}
\end{figure}

\par The fraction of thermal energy that was released, compared to the total energy, was smaller for higher $A_\mathrm{eff}$. The fraction of particles lost during the crash, compared to the total amount of particles, decreased as well from $A_\mathrm{eff} = 2.0$ to $A_\mathrm{eff} = 2.5$, but from $A_\mathrm{eff} = 2.5$ to $A_\mathrm{eff} = 3.0$, it was slightly rising. Table \ref{tab:multi_n_quantities} and figure \ref{fig:elm_parameters} also state the pedestal height reduction of the temperature and density profiles. The pedestal heights where determined by taking the local temperature and density at $\psi_N = 0.9$. The particle density pedestal top losses increase stronger from $A_\mathrm{eff} = 2.5$ to $A_\mathrm{eff} = 3.0$ than the total particle content. This is because the latter also includes the area of the steep gradient ($\psi_N > 0.93$), where the losses for $A_\mathrm{eff} = 3.0$ are smaller compared to those of $A_\mathrm{eff} = 2.5$, as shown in the lower right plot of figure \ref{fig:elm_profiles}. 

\par The decreasing loss of energy contradicts the experimental findings of \cite[13-14]{frassinettiEffectIsotopeMass2023} introduced in section \ref{different_isotopes_fundamentals}, which indicated an increase of the relative energy loss from around 10\% to around 15\% when moving from $A_\mathrm{eff} = 2.0$ to $A_\mathrm{eff} = 3.0$. In \cite{frassinettiEffectIsotopeMass2023}, there was also no clear change of the relative loss of the temperature pedestal height within measurement uncertainties, when changing the ion mass. But the density pedestal clearly increased for higher ion masses, which is again contrary to the slight reduction or constant behavior seen here. This indicates, that the responsible effects were not captured in the performed simulations. The pedestal in the experiment differed among ion masses, which means that a modification of the input profiles, such as the transport coefficients, would be required for a more direct comparison.

\par Table \ref{tab:multi_n_quantities} and figure \ref{fig:elm_parameters} also show the time when the first unstable mode starts to grow, as well as the maximum of the mode energy (which was determined to correspond to the maximum of thermal energy loss). The difference between start and maximum time is growing very moderately. The start and end of the growth section of the \acrshort{elm} shifts distinctly towards earlier times for higher ion mass. This follows the behavior of the $n=3$ mode, which becomes unstable earlier for higher ion masses.

\clearemptydoublepage
\chapter{Conclusion}
\par As part of this thesis, isotope effects for pedestal instabilities have been studied based on an ASDEX Upgrade H-Mode scenario. The aim is to contribute to a better understanding of edge localized mode dynamics under different plasma compositions, while removing any differences from the intrinsic changes of turbulent and neoclassical transport that result in different pedestal structures with different main ion masses. This has received interest due to recent JET experiments with deuterium, deuterium-tritium and tritium plasmas, which show pronounced differences. In this thesis, it is possible to keep all other plasma properties virtually identical while only the ion mass is modified. This way, a clear isolation of relevant effects can be studied. 

\par In the thesis, concretely, pedestal studies for \acrshort{aug} were performed with single toroidal harmonics to assess the linear stability. Already here, more physics are included than in typical linear MHD studies, since a visco-resistive MHD model is used in fully realistic geometry including core and scrape-off layer region. Furthermore, the self-consistently establishing plasma flows in the MHD model with diamagnetic drift extensions are included, which have a very important effect in the stability of peeling-ballooning modes that underlie ELM stability. The simulations revealed, that the isotope effects due to MHD affect the growth rate of modes near the separatrix, which are known to cause \acrshort{elm}s. This also influences the conditions under which the aforementioned modes become unstable. Altogether, the differences between isotopes on the edge stability were small compared to the overall changes during the build-up of the pedestal and ExB flows within time. Furthermore, it was shown that the ion mass effects originate from MHD's momentum equation. This influences the growth rates predominantly through an isotope dependence when applied perturbatively on the unstable mode, leading to differences in the modes' evolution and thus determining the conditions on the profiles for instability. The ion mass was also shown to affect the magneto-hydrodynamic evolution of the profiles itself, such as pedestal height and ExB flows.

\par Finally, also some non-linear simulations with multiple toroidal harmonics were performed for the violent non-linear phase of an ELM-like event with losses that are in the order of type-I ELMs in the experiment. This was done for three different main ion mass numbers of 2.0, 2.5 and 3.0. The expelled thermal energy and particle content did only differ in the order of 10\% between the ion masses, which is much less compared to experimental results. The variations among ion masses in the lowering of the pedestal top during the ELM crash were small as well. A slight decrease in the density pedestal top's reduction, as well as in the expelled thermal energy was seen for higher ion masses, which is contrary to a strong increase seen of those values in the experiment. This indicates, that other effects not arriving from MHD could be behind the experimental differences. 

\par To further investigate the physics of interest, the non-linear simulations should be extended to capture at least a full \acrshort{elm} cycle, so that the isotope's influence due to MHD on important parameters such as the repetition frequency can be assessed. Also, to move further towards an explanation of the experimental differences, it should be considered to adapt the input profiles so that similar pedestal structures and build-up times as in the experiment are achieved. This would be a first step to assess the influence of effects not captured by MHD, such as small-scale turbulence, on \acrshort{elm}s.

\clearemptydoublepage

\printbibliography[heading=bibintoc]

\clearemptydoublepage
\chapter*{Acknowledgments}
\addcontentsline{toc}{chapter}{Acknowledgments} 
\par This thesis would not have been possible with the support and help of my supervisor, Andres Cathey. His explanations and his patience were vital in enabling me to understand fusion plasmas, MHD and JOREK. Without his guidance, I would not have arrived at the results presented here. Also, his comments were very helpful to sharpen and present my findings, as well as in linking it to previous works. I am indebted to him also because of his very thorough, constructive review of my descriptions. 

\par I am also deeply grateful to Matthias Hölzl, for allowing me to spend time in his research group, as well as for providing further comments and suggestions, which especially helped to place this thesis in a wider context and make it more accessible.

\par During my time at IPP, I also had the pleasure to meet many more interesting people, allowing me to learn more about a physicists' way of thought. I am especially thankful for the openness of the JOREK people in room 130 and the surrounding.

\par Lastly, I want to thank my family for their constant support and encouragement when navigating through life.

\end{spacing}

\end{document}